\documentclass[useAMS,usenatbib]{mn2e}

\title[Feeding vs. Feedback in NGC\,4151. I]{Feeding versus Feedback in  NGC\,4151 probed with Gemini NIFS. I. Excitation}

\author[Storchi-Bergmann et al.]
  {T. Storchi-Bergmann$^1$, P. J. McGregor$^2$, Rogemar A. Riffel$^1$, R. Sim\~oes Lopes$^1$, 
  \newauthor 
   T. Beck$^3$ 
  and M. Dopita$^2$ \\
  $^1$Instituto de F\`\i sica, Universidade Federal do Rio Grande do Sul, Av. Bento Gon\c calves 9500, 91501-970 Porto Alegre RS, Brazil\\
  $^2$Research School of Astronomy and Astrophysics, Australian National University, Cotter Road, Weston Creek, ACT\,2611, Australia\\
  $^3$Gemini Observatory and Space Telescope Science Institute, 3700 San Martin Dr.,  Baltimore, MD\,21218}
\date{Released 2008}

\pagerange{\pageref{firstpage}--\pageref{lastpage}} \pubyear{2007}

\def\LaTeX{L\kern-.36em\raise.3ex\hbox{a}\kern-.15em
    T\kern-.1667em\lower.7ex\hbox{E}\kern-.125emX}

\usepackage{graphicx}

\begin{document}

\label{firstpage}

\maketitle

\begin{abstract}

We have used the Gemini Near-infrared Integral Field Spectrograph (NIFS) to map the emission-line intensity distributions and ratios in the Narrow-Line Region (NLR) of the Seyfert galaxy NGC\,4151 in the Z, J, H and K bands at a resolving power $\ge$\,5000, covering the inner  $\approx$ 200\,$\times$\,300\,pc of the galaxy at a spatial resolution of $\approx$8\,pc. We present intensity distributions in 14 emission lines, which show three distinct behaviours. (1) Most of the ionized gas intensity distributions are extended to $\approx$100\,pc from the nucleus along the region covered by  the known biconical outflow (position angle PA=60/240$\degr$ -- NE--SW), consistent with an origin in the outflow; while the recombination lines show intensity profiles which decrease  with distance $r$ from the nucleus as $I\,\propto\,r^{-1}$, most of the forbidden lines present a flat intensity profile ($I\,\propto\,r^0$) or even increasing with distance from the nucleus towards the border of the NLR.  (2) The H$_2$ emission lines show completely distinct intensity distributions, which avoid the region of the bicone, extending from $\approx$10\,pc to $\approx$\,60\,pc  from the nucleus approximately along the large scale bar, almost perpendicular to the bicone axis. This morphology supports an origin for the H$_2$-emitting gas in the galaxy plane. (3) The coronal lines show a steep intensity profile, described by $I\,\propto\,r^{-2}$; the emission  is clearly resolved only in the case of [Si\,{\sc vii}], consistent with an origin in the inner NLR. 

Using the line-ratio maps [Fe{\sc\,ii}]\,1.644/1.257 and Pa\,$\beta$/Br\,$\gamma$ we obtain an average reddening of E(B-V)\,$\approx$\,0.5 along the NLR and E(B-V)\,$\ge$\,1 at the nucleus. Our line-ratio map [Fe{\sc\,ii}]\,1.257$\mu$m/[P{\sc\,ii}]\,1.189$\mu$m of the NLR of NGC\,4151 is the first such map of an extragalactic source. Together with the [Fe{\sc\,ii}]/Pa\,$\beta$ map, these line ratios correlate with the radio intensity distribution, mapping the effects of shocks produced by the radio jet on the NLR. These shocks probably release the Fe locked in grains and produce an enhancement of the [Fe{\sc\,ii}] emission at $\approx$\,1\arcsec\ from the nucleus. At these regions, we obtain electron densities N$_e \approx\,4000$\,cm$^{-3}$ and temperatures T$_e\,\approx\,15000$\,K for the [Fe{\sc\,ii}]-emitting gas. For the H$_2$-emitting gas we obtain much lower temperatures of T$_{exc}\,\approx$\,2100\,K and conclude that the gas is in thermal equilibrium. The heating necessary to excite the molecule may be due to X-rays escaping perpendicular to the cone (through the nuclear torus, if there is one) or to shocks probably produced by the accretion flow previously observed along the large scale bar.

The  distinct  intensity distributions and physical properties of the ionized and molecular gas, as well as their locations, the former along the outflowing cone, and the latter in the galaxy plane surrounding the nucleus, suggest that the H$_2$-emitting gas traces the AGN {\it feeding}, while the ionized gas traces its {\it feedback}.

 
\end{abstract}


\section{Introduction}

NGC\,4151 is the nearest and apparently brightest Seyfert 1 galaxy (or Seyfert 1.5, according to \citet{ost76}), and thus harbours one of the best studied active galactic nuclei (hereafter AGN) \citep{crenshaw07}. As pointed out by \citet{mundell99b}, from old optical observations, NGC\,4151 was believed to be a small spiral galaxy with major axis extending by $\approx$2$\farcm$5 along position angle (hereafter PA) PA$\approx$130$\degr$. But this is just the inner part of the galaxy, comprising an oval distortion, or  ``weak fat bar'' \citep{mundell99a}, beyond which there is a much larger disk, weak in the optical, but bright in H{\sc\,i} emission, with spiral arms extending up to 6\,arcmin from the nucleus \citep{davies73}. From the H{\sc\,i} kinematics, it was concluded that the galaxy has a small inclination of $i\approx$21$\degr$, and a major axis PA$\approx$22$\degr$ \citep{pedlar92,davies73,mundell99b,das05}. Its radial velocity is $cz=997$\,km\,s$^{-1}$ \citep{pedlar92}, the Hubble type is (R')SAB(rs)ab, and we will adopt in this paper a distance of 13.3\,Mpc, corresponding to a scale at the galaxy of 65\,pc\,arcsec$^{-1}$ \citep{mundell03}.

In radio continuum observations NGC\,4151 shows a linear structure comprising several knots elongated over $\sim 3\farcs$5 along PA$\approx$77$\degr$, which is embedded in a diffuse emission extending over $\sim 10\farcs$5 \citep{pedlar93,mundell95}. More recent high resolution radio images \citep{mundell03} reveal a faint jet underlying the discrete components which seem to be shocklike features produced by interactions of the jet with gas clouds in the galaxy, as well as neutral gas absorption consistent with being due to an obscuring torus.

In the optical, the narrow-line region (hereafter NLR) of NGC\,4151 has been found to have a biconical morphology (both in [O{\sc\,iii}]$\lambda$5007 and H$\alpha$ emission lines), with the line-of-sight outside but close to the edge of the cones \citep{evans93,hut98}. The projected opening angle of the cones is $\sim$75$\degr$ and the projected axis is oriented along a position angle PA$\sim$60/240$\degr$. Optical spectroscopy reveals outflows along the cones,
with the approaching side to the SW \citep{med95,evans93}. The outflows have been modelled in a number of studies \citep[e.g.][]{das05,crenshaw00,hutchings99}, in which it is also argued that the AGN should be the primary ionization source of the bicone as no clear correlation with the radio jet is observed. The radio jet may be nevertheless associated with high velocity clouds observed in the NLR \citep{winge97}. 

Combining X-ray, UV and optical spectra, \citet{kra05,kra06} and \citet{crenshaw07} where able to characterize outflows estimated to be at only $\sim$0.1\,pc from the nucleus, obtaining a high mass outflow rate ($\sim$0.16\,M$_\odot$\,yr$^{-1}$) which is about 10 times the accretion rate necessary to feed the AGN in NGC\,4151. They sugest that this outflow originates in an accretion disk wind.

In the near-infrared, \citet{thompson95} obtained spectra from 0.87$\mu$m to 2.5$\mu$m, and concluded that, from the high emission-line ratios of [Fe{\sc\,ii}]/H{\sc\,i}, the majority of the Fe should be in gaseous form, thus implying grain destruction to release a significant fraction of the iron usually tied up in dust. He obtained an electron temperature $T_e\approx\,10^4$K and density $N_e\approx\,10^4$\,cm$^{-3}$ for the region emitting  [Fe{\sc\,ii}]  lines. Long-slit J-band observations of the kinematics of the [Fe{\sc\, ii}]$\lambda$1.2570$\mu$m and Pa$\beta$ emission lines showed similar velocity structure to that observed in [O{\sc iii}] \citep{knop96}, which was later confirmed with two-dimensional Integral Field Unit (hereafter IFU) observations \citep{turner02}. This similarity supports the same origin for [O{\sc\,iii}] and [Fe{\sc\,ii}] emission, namely photoionization by the AGN in the NLR. Nevertheless,  broadening of the [Fe{\sc\,ii}] emission relative to Pa$\beta$, which is observed along the NLR \citep{knop96} suggests additional processes contributing to the [Fe{\sc\,ii}] emission, such as shocks from  an AGN wind or jet.

Despite being a well-studied galaxy \citep{ulrich00}, the dynamics and excitation of the NLR, as well as the role of the radio jet, are  not yet fully understood. In the present paper we use the Gemini Near-infrared Integral Field Spectrograph (NIFS) equipped  with the adaptive optics module ALTAIR -- to map the NLR gas distribution and excitation in the inner $\approx 200$\,pc$\times$400\,pc, at a spatial resolution of $\approx$7\,pc at the galaxy. The data cover the wavelength range 0.95--2.51~$\mu$m at a spectral resolving power over 5000. 
In a companion paper \citep{sl08} (hereafter Paper II), we use these data to obtain the gas kinematics and present emission-line channel maps of the NLR obtained by slicing the strongest emission-line profiles in velocity bins of 60\,km\,s$^{-1}$. 

The present paper is organized as follows. In section 2 we describe the observations and reductions. In section 3 we present the results, which include flux measurements of 55 emission lines and 14 intensity distribution maps as well as line ratio maps. In section 4 we derive physical parameters for the NLR and discuss the origin of the [Fe{\sc\,ii}] and H$_2$ emission, and in section 5 we present our conclusions.

\section{Observations and Reductions}
\label{data}

Two-dimensional spectroscopic data were obtained on the Gemini North 
telescope with the NIFS instrument \citep{nifs03} operating with 
the ALTAIR adaptive optics module on the nights of December 12, 13, and 16 
2006 UT. ALTAIR was used in its Natural Guide Star mode with optical light 
from the nucleus of NGC\,4151 feeding the adaptive optics wave front 
sensor. The uncorrected seeing FWHM, as reported by ALTAIR, was generally 
in the range 0.6-0.9\arcsec, measured in  the V-band, but reached 1.2\arcsec\ on some occasions. 
The observations covered the standard $Z$, $J$, $H$, and $K$ spectral 
bands at two-pixel resolving powers of 4990, 6040, 5290, and 5290, 
respectively. This resulted in wavelength coverage of 0.94-1.16 $\mu$m, 
1.14-1.36 $\mu$m, 1.49-1.81 $\mu$m, and 1.99-2.42 $\mu$m, respectively. 
Additional spectra were obtained at the $K_{long}$ setting of the $K$ 
grating. This covers the wavelength range 2.09-2.51 $\mu$m, which includes 
the H$_2$ Q-branch.

NIFS has a square field of view of $\approx$ 3.0\arcsec\ $\times$ 
3.0\arcsec, divided into 29 slitlets each 0.103\arcsec\ wide with a 
spatial sampling of 0.042\arcsec\ along each slitlet. 
The FWHM of the spatial profile of a star is 0$\farcs12\,\pm\,0\farcs$02
at the  H, K and K$_{long}$ bands, corresponding to $\approx$8\,pc at the galaxy, 
while at the J and Z bands it is larger, 0$\farcs16\,\pm\,02$, corresponding to  
$\approx$10\,pc at the galaxy.
This is dominated by the 0$\farcs$1 slitlet width across 
the slitlets and by instrumental aberrations along the slitlets. But we have verified
an increasing strength of  the uncorrected seeing halo towards shorter wavelengths, 
which degrades the image quality. 
In order to gauge the performance of our data in terms of image quality, 
we have measured the flux in a 0$\farcs$2  diameter circular
aperture and in a 1$\farcs$5 diameter circular aperture for each of the telluric standard 
stars. The resulting ratio between the flux in the smaller aperture to the one in the 
larger aperture is 0.19 for the $Z$ band, 0.27 for the $J$ band, 0.30 for the $H$ band, 
0.47 for the $K$ band, and 0.43 for the $K_{long}$ band. 

Fig.\,\ref{psf} shows the spatial profiles of a star in the different bands, as compared with
the spatial profiles in the galactic continuum in the Z, J K and K$_{long}$ bands.
It can be argued that, as the observations were obtained using the galaxy nucleus as a guide star, the real PSF should be that derived from  the spatial profile of the nuclear source, instead of the stellar profile. It can be seen that in the K band, the FWHM of the nuclear source profile
is almost indistinguishable from that of the star, while in the K$_{long}$ band it is 0$\farcs$04 
larger thus 0$\farcs$16. In the J band, the FWHM of the nuclear source is smaller that
of the stellar profile, thus we adopt for the PSF the larger stellar value of 0$\farcs16\,\pm\,02$, while in the Z band the FWHM of the nuclear source is significantly larger than that of the star, 0$\farcs$24. In the H band the FWHM of the nuclear source profile is 0$\farcs15\,\pm\,0\farcs\,02$. Thus a representative value for the FWHM of the PSF in the J, H, K and K$_{long}$ bands is 0$\farcs14\,\pm\,0\farcs\,02$, corresponding to a spatial resolution at the galaxy of 9\,$\pm$\,1.3\,pc. In the Z band the resolution is poorer, corresponding to 15\,$\pm$\,1.3\,pc at the galaxy.

The instrument was set to a position angle of 345 degrees to align 
the slitlets approximately perpendicular to the axis of the radio jet in 
NGC\,4151 \citep{mundell03}. This results in coarser spatial sampling 
along the jet and finer spatial sampling across it. The $Z$, $J$, 
$H$, and $K$ observations covered three adjacent NIFS fields 
centred on the NGC\,4151 nucleus and offset by $\pm$ 2$\farcs$5 along 
position angle 75 degrees. The resulting field of view is 8.0\arcsec\ 
$\times$ 3.0\arcsec. Only a single NIFS field was obtained at the 
$K_{long}$ grating setting due to the limited extent of the H$_2$ emission. 
This was centred on the nucleus at the same position angle as the other 
observations.

Each dataset was recorded as a sequence of two 90 s exposures at each of 
the three field positions on NGC\,4151 followed by two 90 s sky exposures. 
The nucleus did not saturate in this time. The sky positions were 
displaced by $\approx \pm$75\arcsec\ from NGC\,4151 along PA = 75 degrees 
and dithered by $\pm\,0\farcs$2. This sequence was repeated three times 
and concluded with a fourth object set. It resulted in eight object frames 
at each field position and six offset sky frames for each grating setting. 
An arc spectrum was obtained along with each dataset, and spectra of the 
nearby Hipparcos stars HIP\,56324 (A3V) and/or HIP\,61471 (A0V) were 
obtained before and/or after the NGC\,4151 observations to provide telluric 
correction. The Hipparcos star observations were also used for flux 
calibration. 

The data reduction was accomplished using tasks contained in the {\sc nifs}
package, which is part of the {\sc gemini iraf} package, as well as generic {\sc iraf} 
tasks. The reduction procedure first applied a linearity correction and 
then subtracted a median-combined sky frame, multiplied by a flatfield 
frame, and cut each object frame into 29 sub-images, one for each NIFS 
slitlet. Bad pixels identified in the flatfield and dark frames were then 
removed by interpolation. A coordinate transformation was then applied to 
each two-dimensional sub-image to linearize the wavelength and spatial 
scales. These were derived from the arc exposure and exposures of the 
flatfield lamp with a Ronchi grating aligned perpendicular to the NIFS 
slitlets, respectively. The transformed two-dimensional images were then 
stacked into a three-dimensional data cube with two spatial and one 
spectral dimension. The data cubes for each object exposure were then 
collapsed in the spectral direction to produce a continuum image of the 
sky and the centroid of the NGC\,4151 nucleus was measured. The individual 
data cubes were then recentered to remove tracking drift and combined 
using the {\sc iraf imcombine}  task. Each spectrum in the combined data cube was 
then corrected for telluric absorption based on the spectrum of the A-type 
Hipparcos star after intrinsic hydrogen absorption had been removed by 
Gaussian fitting. The spectra were then flux calibrated by reference to 
the telluric-corrected spectrum of the Hipparcos star, which was assumed 
to have a blackbody shape over the wavelength range of each near-infrared 
spectral band and an average absolute flux density defined by the 
appropriate 2MASS $J$, $H$, or $K$ broadband magnitude. The total signal 
of the Hipparcos star was measured in a 1.5\arcsec\ diameter aperture so 
includes as much of the uncorrect adaptive-optics halo light from the star 
as practical to measure in our 3.0\arcsec\ $\times$ 3.0\arcsec\ field. As 
such, the flux calibration applies to $detected$ light per spatial pixel: 
no correction is attempted to the total flux in the 
adaptive-optics-corrected point spread function. 

Inspection of individual spectra of telluric 
stars in the Z and J bands showed flux variations of 18\% 
in data obtained just before and after the galaxy observations. 
These fluctuations indicate that the night of December 13, 2006, 
when these observations were made, had variable seeing, as
there is no report of the presence of clouds during the night.
During the previous night, when the H and K band observations were
made, and on December 16, when the K$_{long}$ observations were made,
the seeing was stable. As a result, the flux-calibrated spectra of the galaxy show a
flux excess of $\approx$17\% when we compare the red end of the J band spectra 
to the blue end of the H spectra. We have thus 
re-calibrated the Z and J band fluxes dividing them by the factor 1.17, in order
to have consistent data over all spectral bands.

The final data cubes contain 2250 spectra per band, with each spectrum 
corresponding to a spatial coverage of $6.6\times2.7$ pc$^2$ at the 
galaxy. Although the total field covered in the $Z$, $J$, $H$, and $K$ 
observations is 3\arcsec\ $\times$ 8\arcsec, most of the line emission, 
which is the subject of the present study, is concentrated within the 
inner 3\arcsec\ $\times$ 5\arcsec, corresponding to a region of dimensions 
192 pc $\times$ 320 pc at the galaxy.

\section{Results}

\begin{figure}
\includegraphics[scale=0.9]{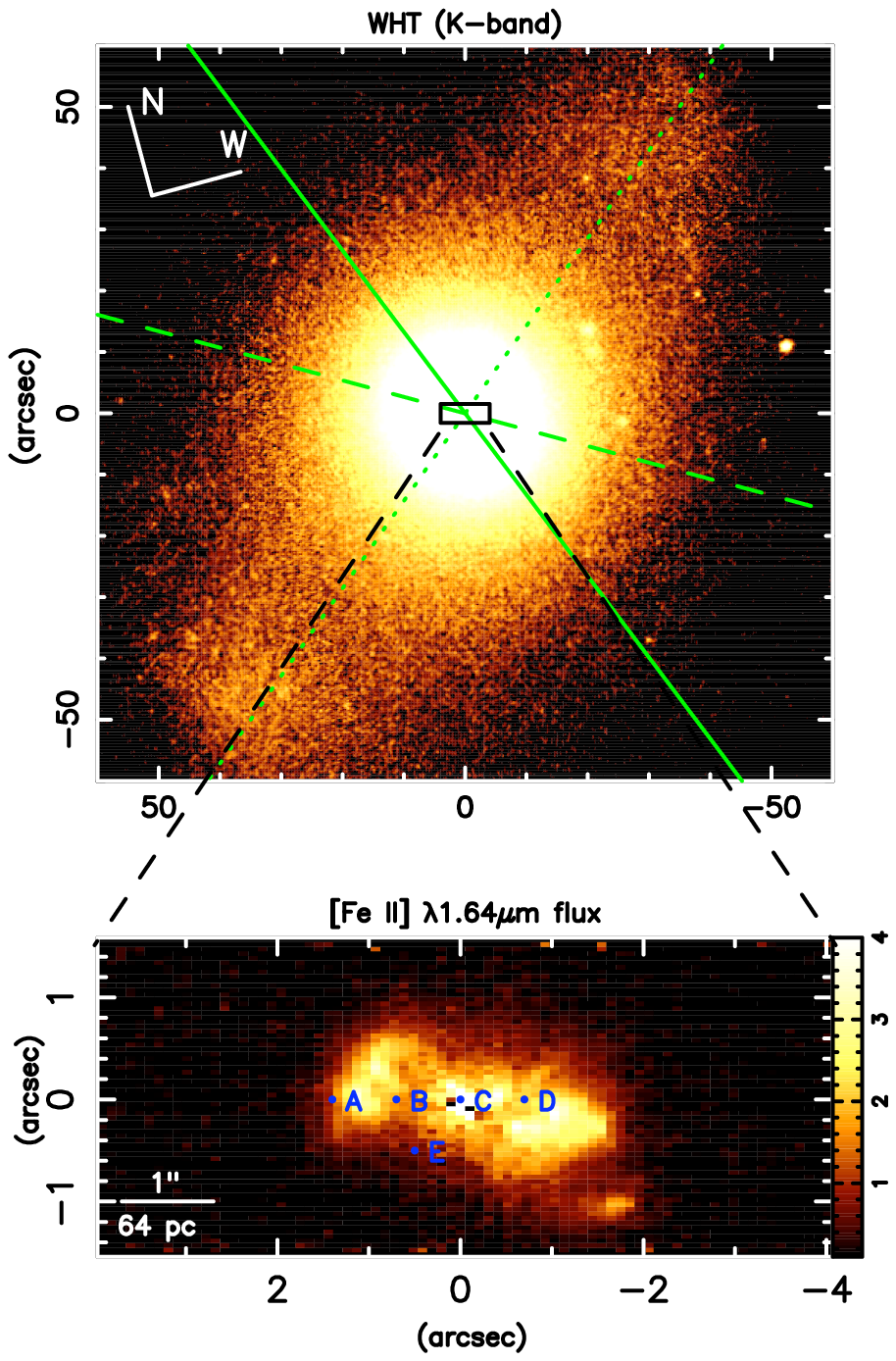}
\caption{Top: K-band image of the central 60\arcsec$\times$60\arcsec of NGC\,4151 obtained with WHT. The image has been rotated to the same orientation of the NIFS frames. The continuous line shows the orientation of the major axis of the galaxy, the dashed line shows the orientation of the bicone and the dot-dashed line shows the orientation of the bar (visible in the image). The rectangle shows the region covered by the NIFS observations. Bottom: image obtained from the NIFS frames integrating in the [Fe{\sc\,ii}]$\lambda\,1.644\mu$m emission line; the letters identify locations corresponding to the spectra shown in Figs.\,\ref{spectra1} and \ref{spectra2}.}
\label{ima}
\end{figure}

\begin{figure*}
\centering
\includegraphics[scale=1.2]{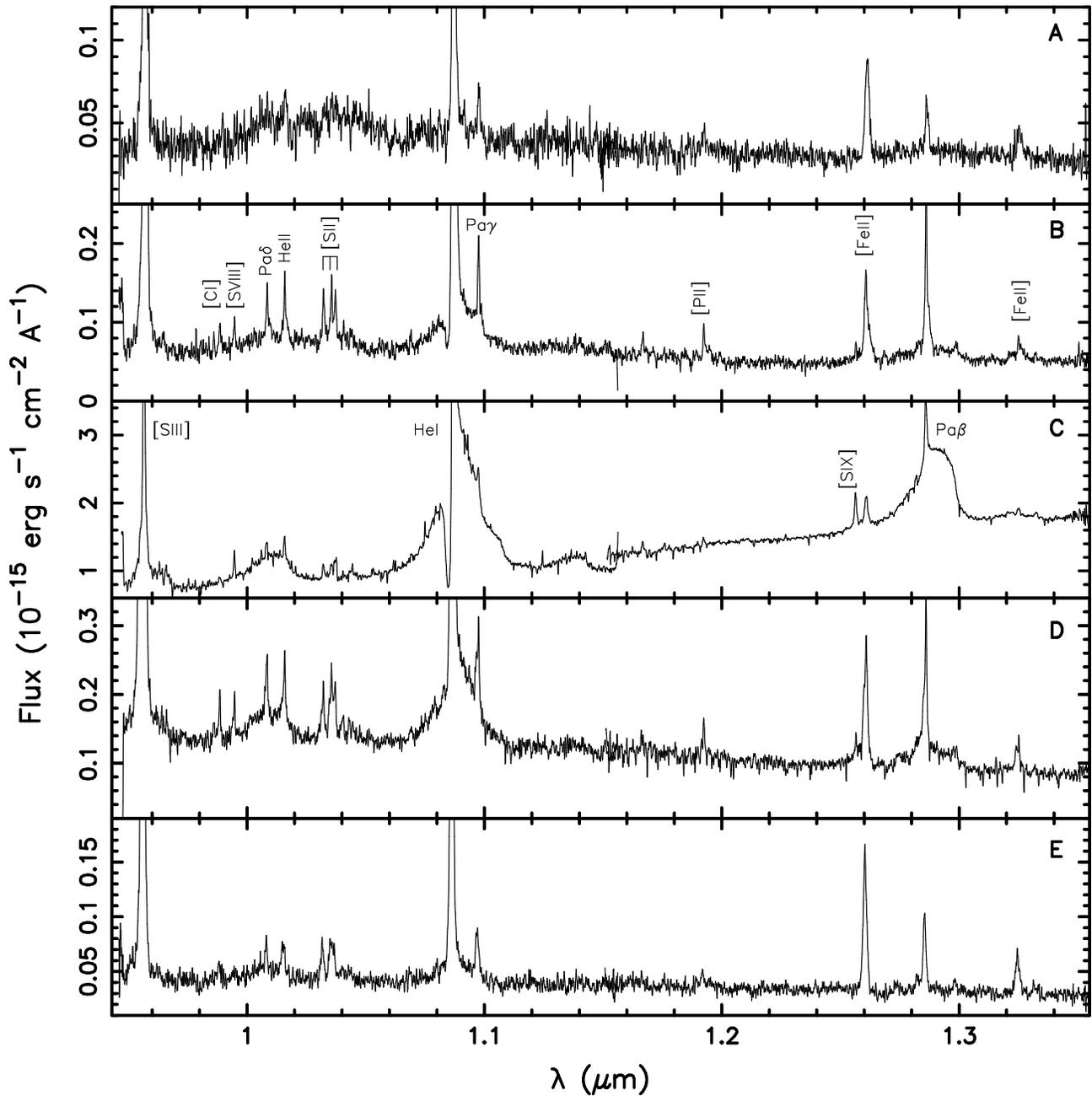}
\caption{Sample of spectra in the Z and J bands, with the corresponding locations identified in the NIFS [Fe{\sc\,ii}] image shown in Fig.\,\ref{ima}. From top to bottom: A: spectrum from a location 1$\farcs$4 E of the nucleus (PA=75$\degr$); B: from 0$\farcs$7 E of the nucleus (PA=75$\degr$); C: spectrum from the nucleus; D: spectrum from 0$\farcs$7 W of the nucleus (PA=255$\degr$); E: spectrum from 0$\farcs$7 SE of the nucleus (PA=120$\degr$), where there is a maximum in the H$_2$ emission (Fig.\,\ref{flux2}).}
\label{spectra1}
\end{figure*}

\begin{figure*}
\centering
\includegraphics[scale=1.2]{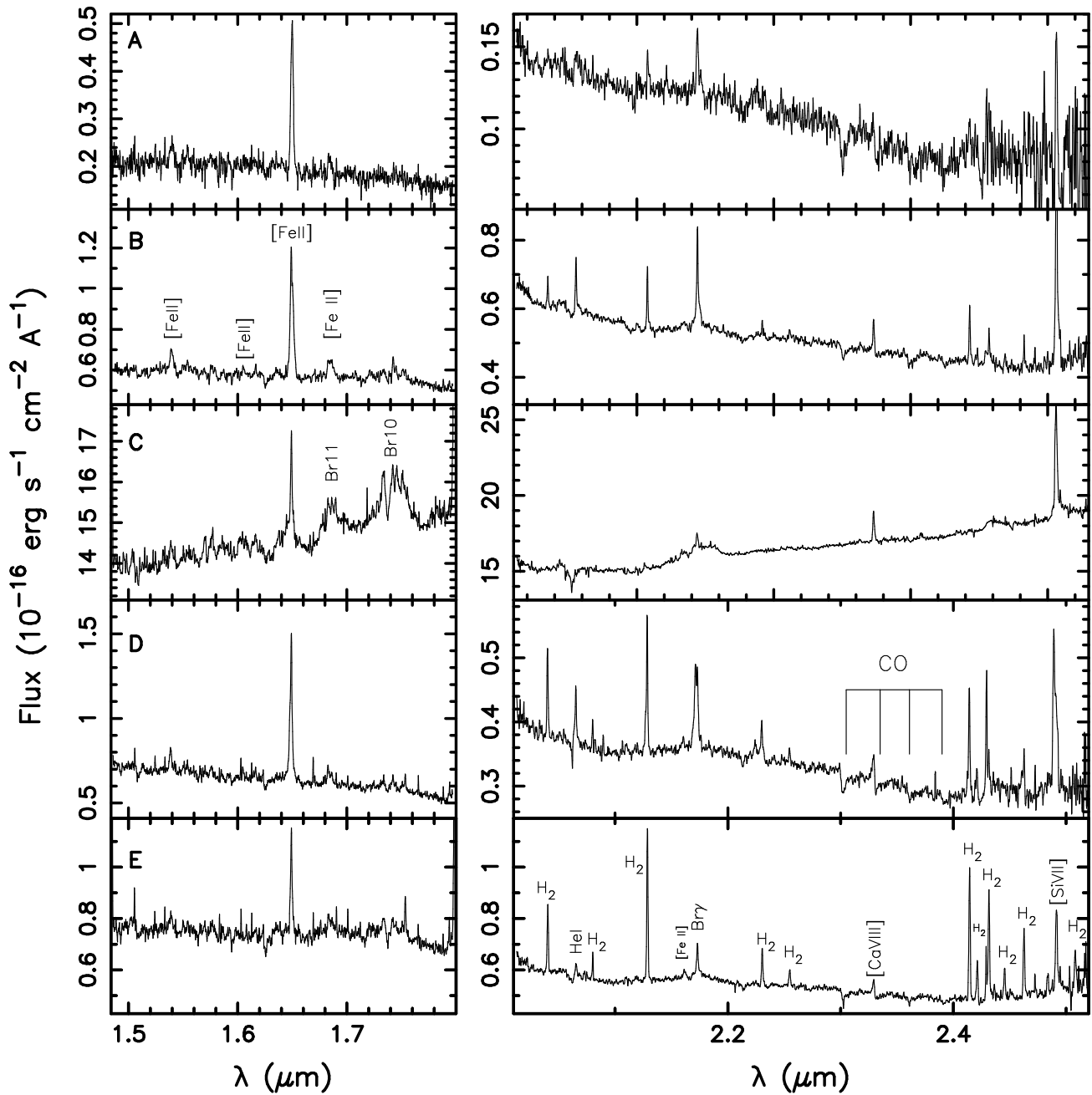}
\caption{Sample of H an K-band spectra, with the corresponding locations identified in the NIFS [Fe{\sc\,ii}] image shown in Fig.\,\ref{ima}, and specified in the caption of Fig.\ref{spectra1}.} 
\label{spectra2}
\end{figure*}

\begin{figure*}
\centering
\includegraphics{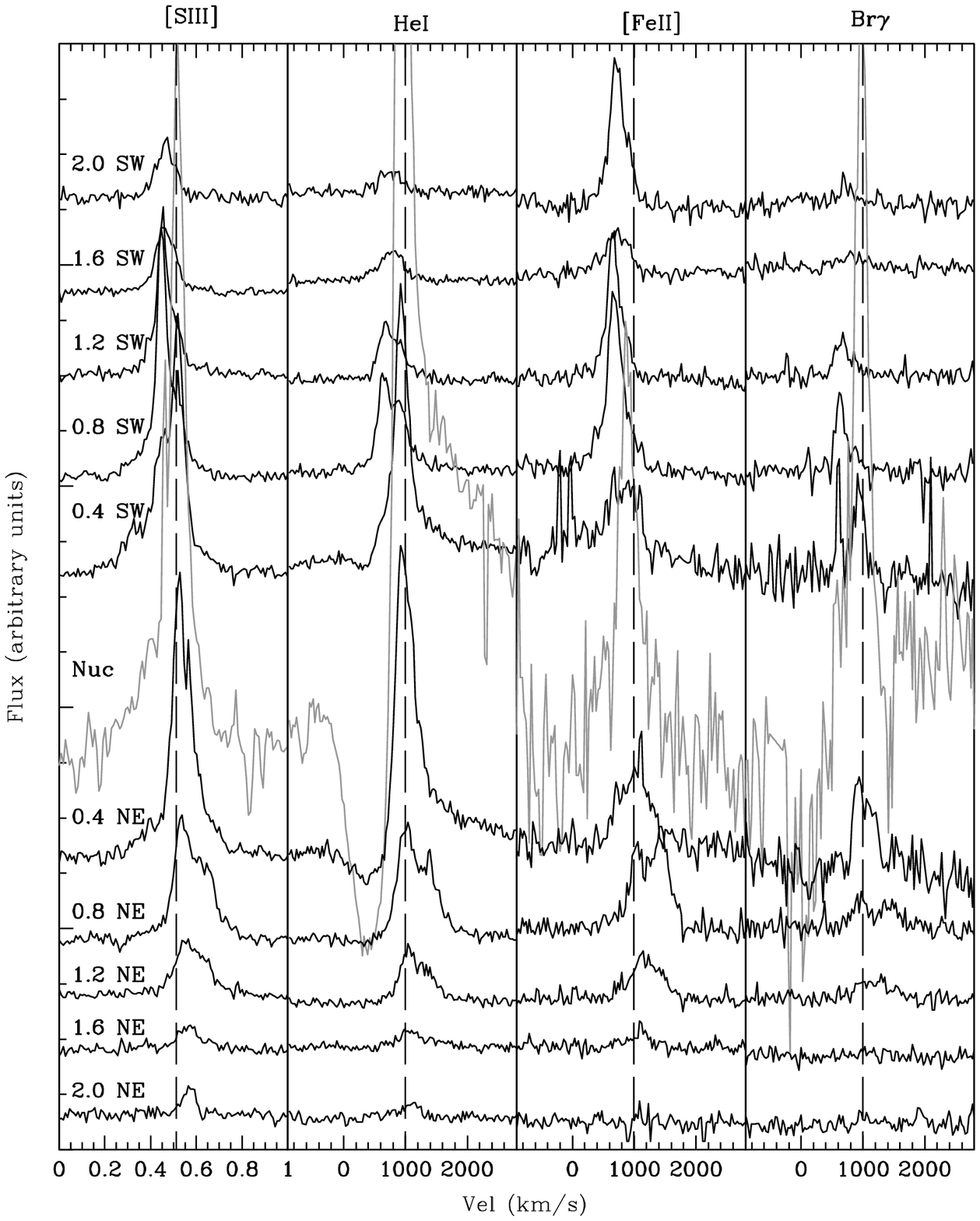}
\caption{Sample of profiles of the [S{\sc\,iii}]\,0.9533$\mu$m, He{\sc\,i}\,1.0833$\mu$m, [Fe{\sc\,ii}]\,1.2570\,$\mu$m and Br\,$\gamma$\,2.1661\,$\mu$m emission lines, from spectra extracted  along the bicone axis (PA=60$\degr$), at the locations relative to the nucleus indicated in the figure. The same flux scale is kept at all locations for each emission line, but varies from line to line.}
\label{profiles}
\end{figure*}

In the top panel of Fig.\,\ref{ima} we present a K-band image of the central 60\arcsec$\times$60\arcsec\ of NGC\,4151, obtained with the William Herschel Telescope (WHT), where the bar can be observed at PA=130$\degr$. We note that the position of the major axis of the galaxy (22$\degr$) is almost perpendicular to the bar.
The central rectangle shows the field-of-view covered by the NIFS observations. In the bottom panel we present an image obtained from the NIFS observations integrating the flux of the [Fe{\sc\,ii}]\,1.6440$\mu$m emission line, where we have marked the positions of representative spectra shown in Figs.\,\ref{spectra1} and \ref{spectra2}.

The spectra shown in Fig.~\ref{spectra1} cover the Z and J bands, while those in Fig.~\ref{spectra2} cover the H and K bands. 
From top to bottom we show spectra from locations approximately at 1$\farcs$4 and 0$\farcs$7 E of the nucleus (PA=75$\degr$), from the nucleus, from 0$\farcs$7 W of the nucleus (PA=255$\degr$), and from 0$\farcs$7 SE of the nucleus (PA=120$\degr$), respectively. The emission lines identified in the spectra are listed in Table\,\ref{fluxes}, together with the corresponding fluxes at the nucleus, at   0$\farcs$9 SW ($\farcs$2 to the right of position D in Fig.\,\ref{ima}, where there is a peak in the [Fe{\sc\,ii}] emission), 0$\farcs$7 SE of the nucleus (position E in Fig.\,\ref{ima}, where there is a peak in the H$_2$ emission) and 0$\farcs$7 E (position B in Fig.\,\ref{ima}), for an aperture of 0$\farcs 31\times 0\farcs 31$ (obtained by binning 3 pixels along the x-axis and 7 pixels along the y-axis).

\begin{table*}
\begin{scriptsize}
\caption{Emission line fluxes within 0$\farcs3 \times 0\farcs 3$ apertures (10$^{-15}$ erg\,cm$^{-2}$\,s$^{-1}$)}
\centering
\begin{tabular}{l l c c c c}
\hline
$\lambda$ (vac.) & ID           	             &   Nucleus   & [Fe\,{\sc ii}] peak (0$\farcs$9\,SW) &  H$_2$ peak (0$\farcs$7\,SE)  & Pos. B (0$\farcs$7\,E) \\
\hline

 0.95486  & H\,{\sc i}\,Pa\,$\epsilon$ (broad)$^\dagger$         &      87.91  $\pm$    11.5   &        -            &             -           &         -       \\
 0.95332  & [S\,{\sc iii}]\,$^1D_2-^3P_2$$^\dagger$                  &    108.57  $\pm$   17.7   & 42.91 $\pm$ 0.18    &    12.83  $\pm$     0.09 &  38.63 $\pm$  0.14  \\
 0.98268  & [C\,{\sc i}]\,$^1D_2-^3P_1$               &       0.29  $\pm$    0.16   &  0.22 $\pm$ 0.08    &    0.09  $\pm$     0.05 &  0.40  $\pm$  0.10  \\
 0.98530  & [C\,{\sc i}]\,$^1D_2-^3P_2$             &       1.44  $\pm$    0.39   &  0.84 $\pm$ 0.18    &    0.46  $\pm$     0.09 &  0.59  $\pm$  0.09  \\
 0.99154  & [S\,{\sc viii}]\,$^2P^0_{1/2}-^2P^0_{3/2}$         &       4.42  $\pm$    0.29   &  0.83 $\pm$ 0.24    &    0.37  $\pm$     0.10 &  0.62  $\pm$  0.10  \\
 1.00521  & H\,{\sc i}\,Pa\,$\delta$ (narrow)            &       3.56  $\pm$    0.34   &  1.52 $\pm$ 0.19    &    0.84  $\pm$     0.15 &  1.61  $\pm$  0.18  \\
 1.00521  & H\,{\sc i}\,Pa\,$\delta$ (broad)$^\ddagger$ &  &  &  & \\
 1.01264  & He\,{\sc ii}\,$5-4$  (narrow)               &       4.34  $\pm$    0.26   &  1.52 $\pm$ 0.20    &    0.57  $\pm$     0.10 &  1.48  $\pm$  0.12  \\
 1.01264  & He\,{\sc ii}\,$5-4$ (broad)$^\ddagger$     &    114.56 $\pm$ 19.1 & -  &  - & - \\  
 1.02895  & [S\,{\sc ii}] \,$^2P^0_{3/2}-^2D^0_{3/2}$            &       3.21  $\pm$    0.34   &  0.89 $\pm$ 0.14    &    0.50  $\pm$     0.10 &  1.12  $\pm$  0.14   \\
 1.03233  & [S\,{\sc ii}]\,$^2P^0_{3/2}-^2D^0_{5/2}$           &       2.72  $\pm$    0.32   &  1.02 $\pm$ 0.13    &    0.44  $\pm$     0.10 &  1.16  $\pm$  0.11   \\
 1.03392  & [S\,{\sc ii}]\,$^2P^0_{1/2}-^2D^0_{3/2}$            &       4.47  $\pm$    0.44   &  1.31 $\pm$ 0.21    &    0.28  $\pm$     0.10 &  1.06  $\pm$  0.14   \\
 1.03733  & [S\,{\sc ii}]\,$^2P^0_{1/2}-^2D^0_{5/2}$            &       1.36  $\pm$    0.26   &  0.31 $\pm$ 0.13    &    0.21  $\pm$     0.07 &  0.50  $\pm$  0.15   \\
 1.04006  & [N\,{\sc i}]\,$^2P^0_{3/2}-^2D^0_{5/2}$             &       1.78  $\pm$    0.37   &  0.20 $\pm$ 0.13    &    0.19  $\pm$     0.12 &  0.26  $\pm$  0.08   \\
 1.04100  & [N\,{\sc i}]\,$^2P^0_{1/2}-^2D^0_{3/2}$             &       1.45  $\pm$    0.31   &  0.07 $\pm$ 0.06    &    0.18  $\pm$     0.10 &  0.44  $\pm$  0.13   \\
 1.06706   & He\,{\sc i}\,$^3S-^3P^0$                      &       1.69  $\pm$    0.43   &  0.13 $\pm$ 0.07    &    0.04  $\pm$     0.04 &  0.32  $\pm$  0.08   \\
 1.08332  & He\,{\sc i}\,$^3P^0-^3S$ (narrow)       &     90.48  $\pm$  3.21   & 22.95 $\pm$ 0.21    &   8.99  $\pm$     0.11 &  30.36  $\pm$  0.18  \\
 1.08332 & He\,{\sc i}\,$^3P^0-^3S$  (broad) $^\star$            &      409.06 $\pm$ 13.29 & - & - & - \\    
 1.09411  & H\,{\sc i}\,Pa\,$\gamma$ (narrow)            &       7.01  $\pm$    0.37   &  2.35 $\pm$ 0.19    &    0.95  $\pm$     0.12 &  1.80  $\pm$  0.11   \\
 1.09411  & H\,{\sc i}\,Pa\,$\gamma$ (broad) $^\star$ & - & - & - & - \\
 1.12900  & O\,{\sc i}\,$^3D^0-^3P$                &       0.53  $\pm$    0.26   &        -            &             -           &         -            \\
 1.16296  & He\,{\sc ii}\,$7-5$                 &       2.65  $\pm$    0.42   &  0.54 $\pm$ 0.17    &    0.31  $\pm$     0.12 &  0.53  $\pm$  0.15   \\
 1.18861  & [P\,{\sc ii}]\,$^1D_2-^3P_2$             &       2.49  $\pm$    0.33   &  1.06 $\pm$ 0.17    &    0.28  $\pm$     0.10 &  0.63  $\pm$  0.10   \\
 1.25235   & [S\,{\sc ix}]\,$^3P_1-^3P_2$             &      8.74  $\pm$    0.38   &  0.52 $\pm$ 0.20    &    0.29  $\pm$     0.10 &  0.39  $\pm$  0.10   \\
 1.25702  & [Fe\,{\sc ii}]\,a$^4D_{7/2}-a^6D_{9/2}$     &       6.41  $\pm$    1.28   &  7.36 $\pm$ 0.18    &    1.44  $\pm$     0.12 &  4.09  $\pm$  0.26   \\
 1.27069  & [Fe\,{\sc ii}]\,a$^4D_{1/2}-a^6D_{1/2}$     &              -              &  1.00 $\pm$ 0.32    &             -           &  0.19  $\pm$  0.15   \\
 1.27912  & [Fe\,{\sc ii}]\,a$^4D_{3/2}-a^6D_{3/2}$    &       0.11  $\pm$    0.07   &  0.60 $\pm$ 0.16    &    0.15  $\pm$     0.06 &  0.49  $\pm$  0.20   \\
 1.28216  & H\,{\sc i}\,Pa\,$\beta$ (narrow)     &      17.38  $\pm$    0.38   &  4.15 $\pm$ 0.16    &    1.26  $\pm$     0.08 &  2.61  $\pm$  0.08   \\
 1.28216  & H\,{\sc i}\,Pa\,$\beta$  (broad)     &     278.52  $\pm$   10.86   &        -            &             -           &  3.35  $\pm$  0.32   \\
 1.29462  & [Fe\,{\sc ii}]\,a$^4D_{5/2}-a^6D_{5/2}$     &                -            &  0.78 $\pm$ 0.17    &             -           &  0.07  $\pm$  0.05   \\
 1.29812  & [Fe\,{\sc ii}]\,a$^4D_{3/2}-a^6D_{1/2}$     &                -            &  0.241 $\pm$ 0.02    &             -           &  0.03  $\pm$  0.02   \\
 1.32814  & [Fe\,{\sc ii}]\,a$^4D_{5/2}-a^6D_{3/2}$      &       1.65  $\pm$    0.39   &  1.86 $\pm$ 0.17    &    0.38  $\pm$     0.11 &  0.50  $\pm$  0.12   \\
 1.53389  & [Fe\,{\sc ii}]\,a$^4D_{5/2}-a^4F_{9/2}$     &        2.21$\pm$1.71        &  0.89 $\pm$ 0.22    &             -           &  0.63  $\pm$  0.31   \\
 1.59991  & [Fe\,{\sc ii}]\,a$^4D_{3/2}-a^4F_{7/2}$     &                -            &  0.48 $\pm$ 0.20    &             -           &  0.11  $\pm$  0.09   \\
 1.64117   & H\,{\sc i}\,Br\,12  (total)                &       29.34 $\pm$ 5.3           &        -            &             -           &         -            \\
 1.64400  & [Fe\,{\sc ii}]\,a$^4D_{7/2}-a^4F_{9/2}$     &       6.2  $\pm$  1.2   &  6.11 $\pm$ 0.19    &    1.29  $\pm$     0.12 &  3.27  $\pm$  0.17   \\
 1.66423  & [Fe\,{\sc ii}]\,a$^4D_{1/2}-a^4F_{5/2}$    &                -            &  0.25 $\pm$ 0.17    &    0.17  $\pm$     0.03 &  0.21  $\pm$  0.05   \\
 1.67734  & [Fe\,{\sc ii}]\,a$^4D_{5/2}-a^4F_{7/2}$    &                -            &  0.56 $\pm$ 0.19    &             -           &  0.04  $\pm$  0.03   \\
 1.68111  & H\,{\sc i}\,Br\,11 (total)                 &         32.29 $\pm$ 8.2            &        -            &             -           &         -            \\
 1.73669  & H\,{\sc i}\,Br\,10  (total)                 &         0.86 $\pm$15.2           &        -            &             -           &  0.43  $\pm$  0.17   \\
 1.74801  & H$_2$\,$1-0\,S(7)$                     &                -            &  0.13 $\pm$ 0.12    &    0.26  $\pm$     0.06 &  0.08  $\pm$  0.09   \\
 1.74890   & [Fe\,{\sc ii}]\,a$^4P_{3/2}-a^4D_{7/2}$     &                -            &  0.12 $\pm$ 0.20    &             -           &  0.03  $\pm$  0.02   \\
 2.03376   &  H$_2$\,$1-0\,S(2)$                     &                -            &  0.19 $\pm$ 0.16    &    0.48  $\pm$     0.04 &  0.27  $\pm$  0.06   \\
 2.05869  &  He\,{\sc\,i}\,$^1P^0-^1S$               &                -            &  0.36 $\pm$ 0.15    &    0.17  $\pm$     0.06 &  0.43  $\pm$  0.07   \\
 2.07498   &  H$_2$\,$2-1\,S(3)$                     &                -            &        -            &    0.13  $\pm$     0.02 &  0.07  $\pm$  0.02   \\
 2.12183  & H$_2$\,$1-0\,S(1)$                      &                -            &  0.42 $\pm$ 0.15    &    1.21  $\pm$     0.03 &  0.55  $\pm$  0.04   \\
 2.15420  & H$_2$\,$2-1\,S(2)$                      &                -            &        -            &    0.09  $\pm$     0.05 &  0.04  $\pm$  0.04   \\
 2.16612  & H\,{\sc i}\,Br\,$\gamma$ (narrow)  &       5.22  $\pm$    0.40   &  0.85 $\pm$ 0.13    &    0.34  $\pm$     0.04 &  0.80  $\pm$  0.05   \\
 2.16612  & H\,{\sc i}\,Br\,$\gamma$ (broad)   &       39.34 $\pm$    3.25   &        -            &             -           &         -            \\
 2.20133   & H$_2$\,$3-2\,S(3)$                      &                -            &        -            &    0.01  $\pm$     0.01 &  0.02  $\pm$  0.02   \\
 2.22344   & H$_2$\,$1-0\,S(0)$                      &                -            &  0.27 $\pm$ 0.13    &    0.28  $\pm$     0.02 &  0.13  $\pm$  0.03   \\
 2.24776  & H$_2$\,$2-1\,S(1)$                      &                -            &        -            &    0.14  $\pm$     0.02 &  0.08  $\pm$  0.04   \\
 2.32204   & [Ca\,{\sc viii}]\,$^2P^0_{3/2}-^2P^0_{1/2}$     &       3.81  $\pm$    0.42   &  0.15 $\pm$ 0.08    &    0.13  $\pm$     0.01 &  0.34  $\pm$  0.03   \\
 2.40847   & H$_2$\,$1-0\,Q(1)$                      &                -            &  0.27 $\pm$ 0.07    &    1.15  $\pm$     0.01 &  0.51  $\pm$  0.01    \\
 2.41367   & H$_2$\,$1-0\,Q(2)$                      &                -            &  0.08 $\pm$ 0.06    &    0.33  $\pm$     0.01 &  0.12  $\pm$  0.01    \\
 2.42180   & H$_2$\,$1-0\,Q(3)$                    &                -            &  0.63 $\pm$ 0.08    &    0.49  $\pm$     0.01 &  0.17  $\pm$  0.01    \\
 2.43697   & H$_2$\,$1-0\,Q(4)$                      &                -            &        -            &    0.33  $\pm$     0.01 &  0.21  $\pm$  0.01    \\
 2.45485   & H$_2$\,$1-0\,Q(5)$                      &                -            &        -            &    0.74  $\pm$     0.01 &  0.25  $\pm$  0.01    \\
 2.47555   & H$_2$\,$1-0\,Q(6)$                     &                -            &        -            &    0.21  $\pm$     0.01 &  0.13  $\pm$  0.01    \\
 2.48334  & [Si\,{\sc vii}]\,$^3P_1-^3P_2$        &      14.29  $\pm$    0.39   &  1.21 $\pm$ 0.07    &    1.17  $\pm$     0.02 &  2.68  $\pm$  0.01    \\
 2.50007   & H$_2$\,$1-0\,Q(7)$                      &                -            &        -            &    0.49  $\pm$     0.01 &  0.20  $\pm$  0.01    \\
\hline     
\multicolumn{5}{l}{$^\dagger$ Narrow component of P\,$\epsilon$ blended with [S\,{\sc iii}]$\lambda$\,0.9533} \\
\multicolumn{5}{l}{$^\ddagger$  Broad component of Pa\,$\delta$ blended with broad component of He\,{\sc i}\,$\lambda$\,1.01264} \\
\multicolumn{5}{l}{$^\star$ Broad component of Pa\,$\gamma$ blended with He{\sc\,i}\,$\lambda$1.08332} \\                                                                              
\end{tabular}                                       
\label{fluxes}                                      
\end{scriptsize}                                    
\end{table*}

Figs.\,\ref{spectra1} and \ref{spectra2} show that the nuclear continuum (panel C) is very red, while the continuum from extra-nuclear regions is blue. An analysis of this continuum, and the constraints which can be derived for the structure producing it -- possibly a dusty torus -- will be presented in a forthcomig paper \citep{riffel09}. The nuclear spectrum shows broad H{\sc\,i} and He{\sc\,i} emission lines, and a deep absorption which is clearly observed on top of the broad profiles in the nuclear He{\sc\,i}\,1.08332$\mu$m, Br$\gamma$ and He{\sc\,i}\,2.05869$\mu$m emission lines. It can also be observed  that there is no H$_2$ emission at the nucleus and that the coronal lines of  [Ca{\sc\,viii}]  and [Si{\sc\,vii}] seem to be extended. The [P{\sc\,ii}]\,1.1886$\mu$m emission line is also extended and can be observed up to 1$\farcs$4 E of the nucleus. This line was first reported in an extragalactic source by \citet{oliva01}, and since then observed in the nuclear spectrum of less than 15 galaxies, including NGC\,4151 \citep{riffel06a}. Combined with  [Fe{\sc\,ii}]1.2570$\mu$m, the [P{\sc\,ii}]  line is a powerful diagnostic of the origin of the [Fe{\sc\,ii}] emission in galaxies \citep{oliva01,jackson07}. In the present paper we provide the first 2D map of an extragalactic source in this emission line.

The profiles of most of the emission lines vary according to the location in the NLR, and an illustration of this variation is shown in Fig.\,\ref{profiles}, where we present a sequence of selected emission-line profiles, from different locations along the bicone (PA=60$\degr$). A vertical dashed line shows the adopted systemic velocity, of 997\,km\,s$^{-1}$ \citep{pedlar92}. 
The same flux scale is kept for each emission line in Fig.\ref{profiles}, so that the flux variation can also be observed as a function of distance from the nucleus. The scale nevertheless varies for the different emission lines. It can be observed that the center of the emission lines drifts from blueshifts observed to the SW of the nucleus to redshifts observed to the NE of the nucleus. In some locations -- e.g., at 0$\farcs$8\,SW of the nucleus in Fig.\ref{profiles} -- many emission lines are clearly double-peaked, while in other locations the profiles are asymmetric suggesting the presence also of two components, although unresolved. The blueshifted absorption in the He{\sc\,i}\,1.08332$\mu$m emission-line profile is observed at  $-468$\,km\,s$^{-1}$ relative to the systemic velocity (more details will be given in Paper II). A similar absorption (with similar blueshift) is observed in Br$\gamma$ and in the He{\sc\,i}\,2.0587$\mu$m lines. The absorption in both He{\sc\,i} lines reach below the interpolated continuum indicating that the absorber covers much of the continuum source and the BLR clouds.  To our knowledge, these absorption features have not been seen in previous near-infrared spectra, such as the one obtained by \citet{ost90} in 1988 or by \citet{thompson95} in 1993.  More recently, \citet{riffel06a} also fail to detect this absorption. Nevertheless, absorptions of similar widths and blueshifts have been observed in the UV and optical.  \citet{kra01} report  absorptions at a velocity $-490$\,km\,s$^{-1}$ with respect to systemic in  HST UV spectra of the NLR of NGC\,4151, which had been previously identified as due to components called {\it D} and {\it E} in a previous study by  \citet{weymann97}, while \citet{hut02} report an absorption  due to He{\sc\,i}\,$\lambda$3889\AA\ at  $-460$\,km\,s$^{-1}$. The absorptions we see in the near-IR are probably related to these UV and optical absorptions seen in HST spectra. Its detection in our spectra is aided by the high spatial resolution of the NIFS data, which minimizes NLR contamination of the nuclear spectrum.

\subsection{Emission-line intensity distributions}

\begin{figure*}
\includegraphics[scale=1.4]{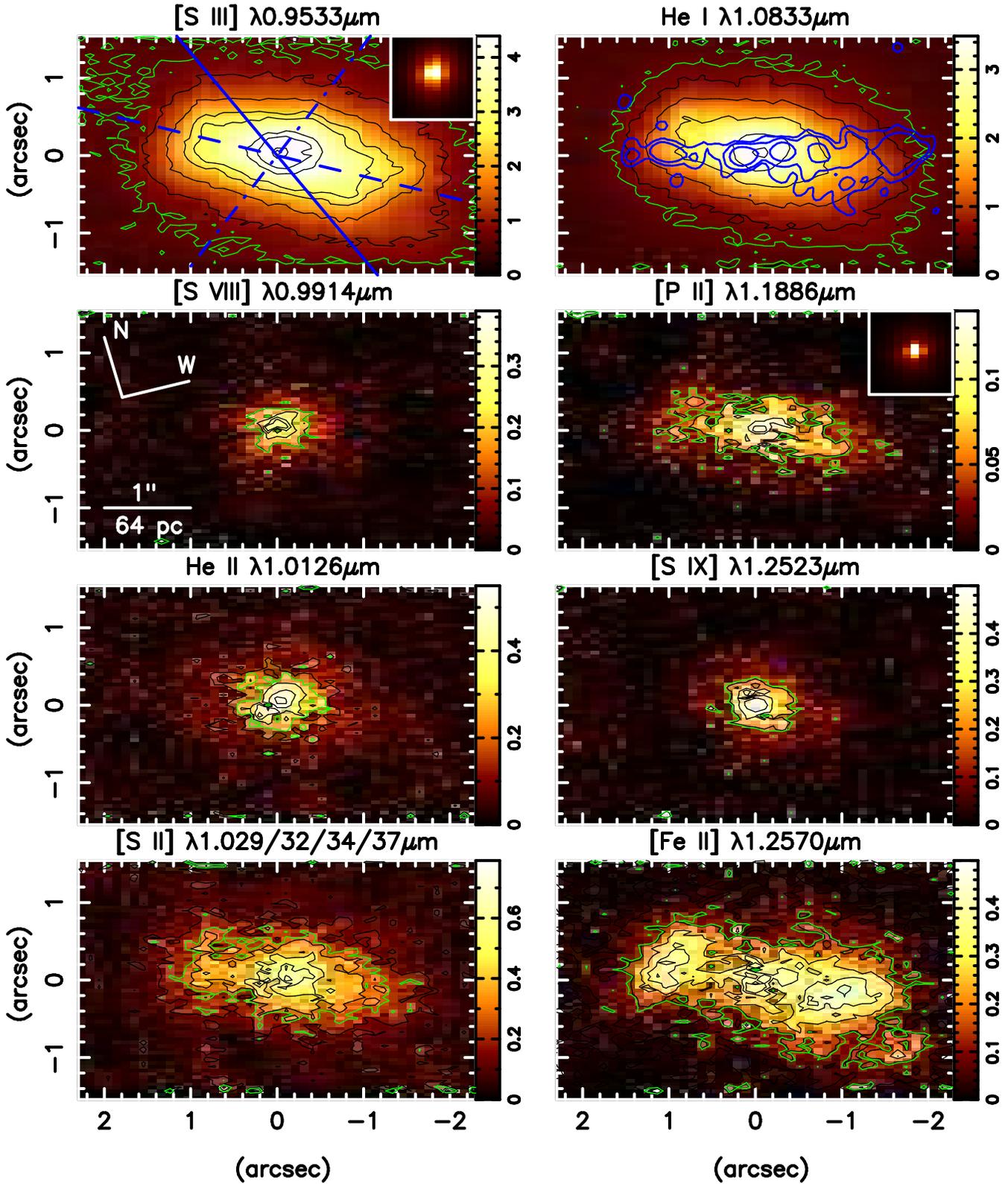}
\caption{Intensity maps of emission lines from the Z and J bands, obtained by integrating the flux under the line profiles after subtraction of the continuum. The green contours correspond to 3 times the background noise (3$\sigma$).  The blue contours are from the radio MERLIN image of \citet{mundell95}. The dashed line on the top left panel shows the orientation of the bicone, the continuous line shows the galaxy major axis, and the dot-dashed line shows the orientation of the large scale bar. The insets show images of telluric stars in the Z band (in the [S{\sc\,iii}] panel) and J band (in the [P{\sc\,ii}] panel) with peak intensity normalized to that of the galaxy in the emission line of the corresponding panel.  Flux units are $10^{-15}$\,erg\,cm$^{-2}$\,s$^{-1}$\,spaxel$^{-1}$.}
\label{flux1}
\end{figure*}

\begin{figure*}
\includegraphics[scale=1.4]{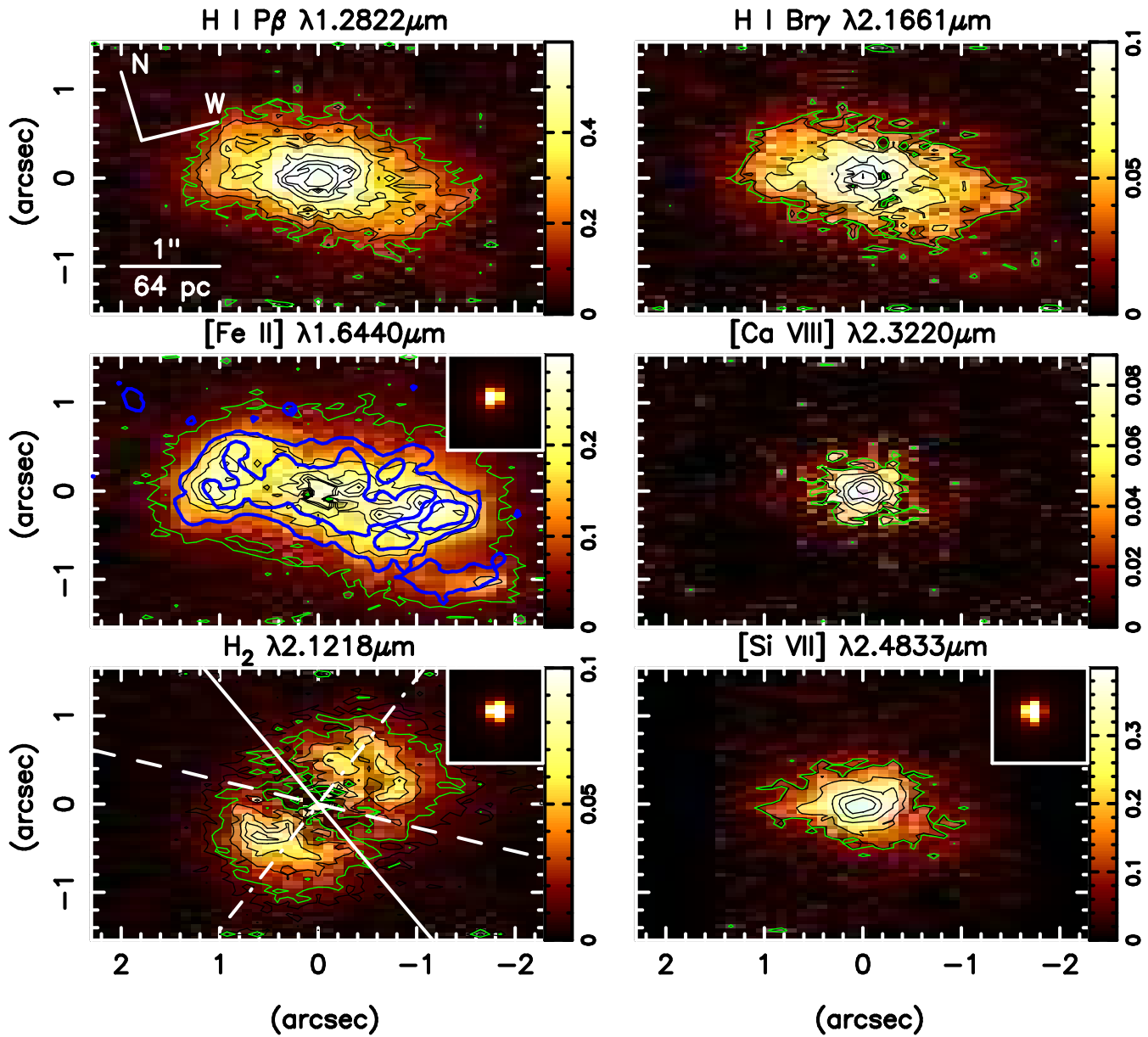}
\caption{Intensity maps of emission lines from the J,  H and K bands, represented as in Fig.\,\ref{flux1}. Blue contours overplotted on the [Fe{\sc\,ii}]$\lambda\,1.644\,\mu$m map are from the [O{\sc\,iii}] image of \citet{hut98}. The insets show images of telluric stars in the H band (in the [Fe{\sc\,ii}] panel), K band (in the H$_2$ panel) and in the K$_{long}$ band (in the  [Si{\sc\,vii}] panel), normalized to the peak intensity of the galaxy in the emission line of the corresponding  panel. }
\label{flux2}
\end{figure*}

\begin{figure*}
\includegraphics[scale=1.4]{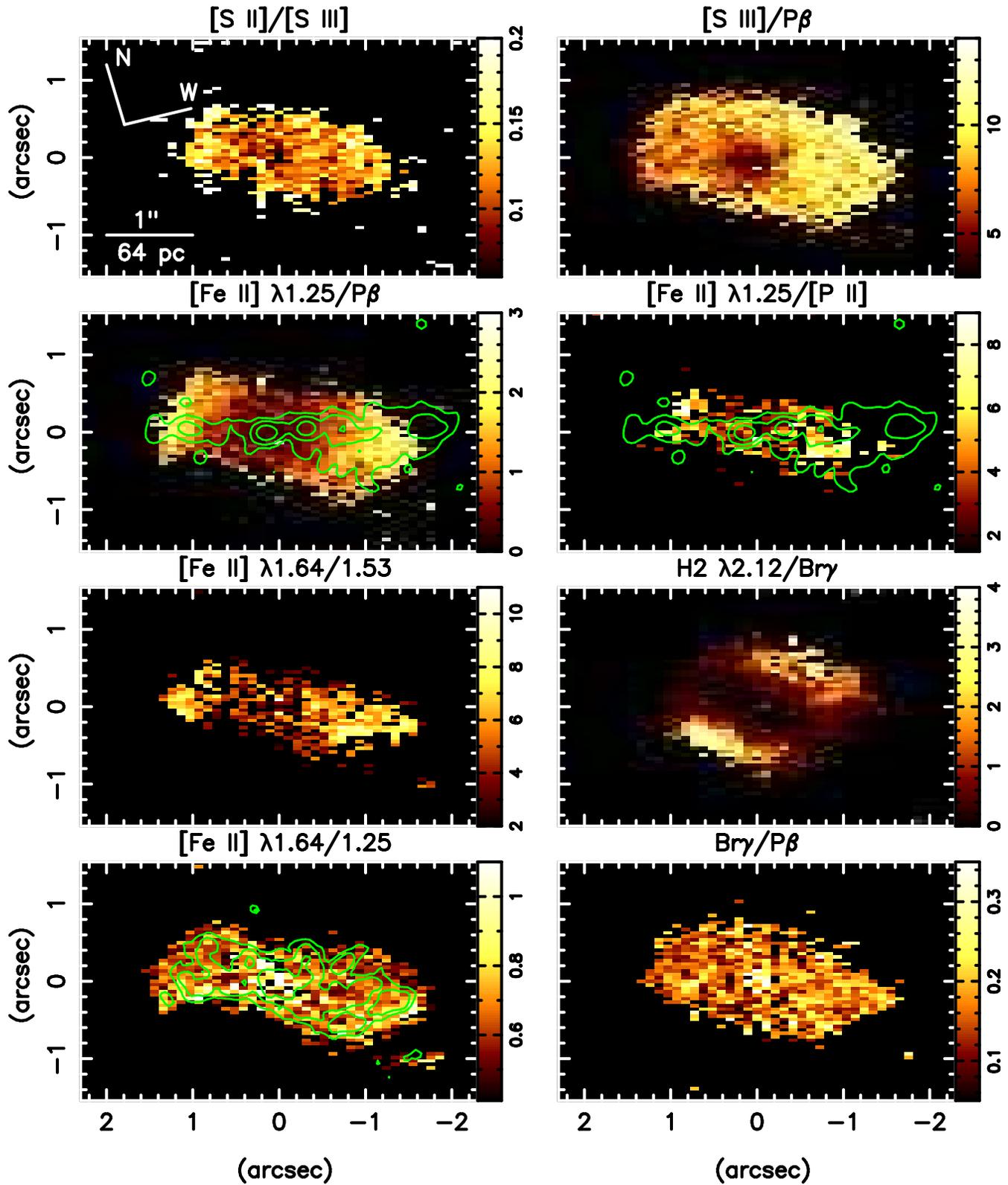}
\caption{Line ratio maps, where the green contours overplotted on the [Fe{\sc\,ii}]/Pa\,$\beta$ map are from the radio MERLIN image of \citet{mundell95}, while the ones overplotted on the [Fe{\sc\,ii}]1.644/1.257 map are from the [O{\sc\,iii}] image of \citet{hut98}.}
\label{ratio1}
\end{figure*}

2D maps of the emission-line intensities have been obtained by integrating the flux under the line profiles, after subtraction of the contribution of the underlying continuum, determined as the average between two spectral windows adjacent to the emission lines. In the case of the H and He lines, which have broad components at the nucleus, the adjacent  continua fall on  top of the broad lines and thus the resulting flux is essentially from the narrow component. Nevertheless, as the broad-line profiles are not symmetric (e.g. Pa$\beta$ profile in the middle panel of Fig.\,\ref{spectra1}) the subtraction is not always perfect and the nuclear fluxes in these lines may have some residual broad-line flux.

The intensity maps for the main extended emission lines from the Z and J bands, namely [S{\sc\,iii}]\,0.9533$\mu$m, [S{\sc\,viii}]\,0.9915\,$\mu$m, He{\sc\,ii}\,1.0126$\mu$m, [S{\sc\,ii}]\,1.029,1.032,1.034,1.037$\mu$m, He{\sc\,i}\,1.0833, [P{\sc\,ii}]\,1.1886$\mu$m,  [S{\sc\,ix}]\,1.2523$\mu$m and [Fe{\sc\,ii}]\,1.2570\,$\mu$m are shown in Fig.\,\ref{flux1}. Along the y-axis, we show the whole extent of the frames (3$\arcsec$), while along the x-axis we show only the region with measurable flux in at least one of the emission lines, comprising the inner 4$\farcs$6. We have overplotted on the He{\sc\,i} flux map the contours (in blue) of a radio $\lambda$21\,cm image from \citet{mundell95}, obtained with MERLIN. These contours have been aligned to the intensity maps under the assumption that component C$_4$ of the radio image is aligned with the peak of our K-band continuum map. Radio component C$_4$ is claimed to contain the active nucleus. We assume that the K-band continuum peak also contains the nucleus. 
The continuous line in the top left panel shows the orientation of the major axis of the galaxy at PA=22$\degr$ \citep{pedlar92}, while the dashed line shows the orientation of the [O{\sc\,iii}] bicone at PA=60/240$\degr$ and the dot-dashed line shows the orientation of the bar, at PA=130$\degr$ \citep{mundell99a}. The [P{\sc\,ii}] flux map shown in the second panel (from top to bottom) to the  right  of Fig.\,\ref{flux1}), is the first 2D map in this line ever obtained for an extragalactic source (as pointed out in the previous section).
The resolution achieved in the images can be judged by a comparison with telluric star images, show as insets in Fig.\,\ref{flux1}: the image of a star in the Z band is shown in the [S{\sc\,iii}] panel, while that in the J band  is shown in the [P{\sc\,ii}] panel, with peak intensity normalized to that of the galaxy in the emission line of the corresponding panel.

In Fig.\,\ref{flux2} we show the intensity maps for additional emission lines observed in the J, H and K bands, namely  Pa\,$\beta$, [Fe{\sc\,ii}]\,1.6440\,$\mu$m, H$_2\,2.1218\,\mu$m, Br\,$\gamma$, [Ca{\sc\,viii}]\,2.3220\,$\mu$m and [Si{\sc\,vii}]\,2.4833\,$\mu$m.  We have overplotted (in blue) on the  [Fe{\sc\,ii}] map  the contours of the [O{\sc\,iii}] emission-line map. 

From Figs. \ref{flux1} and \ref{flux2}, it can be observed that the light distribution varies for different emission lines. For the coronal lines [S{\sc\,viii}], [S{\sc\,ix}] and [Ca{\sc\,viii}], the light distribution is compact and not clearly resolved, 
while the coronal [Si\,{\sc\,vii}]  emission distribution is more extended, being aligned with the bicone to the SW and with the radio jet to the NE. 

Except for H$_2$, the other emission lines show much stronger extended emission, reaching at least 4$\arcsec$ along the bicone and 2$\arcsec$ in the perpendicular direction, corresponding to  projected distances at the galaxy of 256\,pc and 128\,pc, respectively. The intensity distributions are somewhat brighter and more extended to the SW (the near side of the bicone) than to the NE (the far side). 

The  [O{\sc\,iii}] image, whose contours are overplotted on the [Fe{\sc\,ii}] intensity distribution in Fig.\,\ref{flux2} has a ``knotty'' appearance, similar to what is observed in our [Fe{\sc\,ii}] map, with some knots coinciding in both maps. Nevertheless, a detailed comparison shows that the decrease in emission just outside the nucleus and increase again in two opposite regions at about 1$\arcsec$ from the nucleus along the bicone observed in the [Fe{\sc\,ii}] maps is not observed in the [O{\sc\,iii}] map, which shows emission in ``curved strands'' all the way along the bicone \citep{hut98}.

A completely different light distribution is observed in the H$_2$ flux map (bottom left panel of Fig.\,\ref{flux2}): instead of being elongated along the bicone axis, it is elongated almost perpendicular to it. There is  almost no H$_2$ emission within $\approx$0$\farcs$3 from the nucleus, its flux being  distributed in two structures resembling  double arcs  to the NW and  SE, extending from $\approx$0$\farcs$3 to $\approx$1\arcsec from the nucleus along the minor axis of the galaxy (PA=112/292$\degr$), which is also approximately the orientation of the  bar.

In order to look for a possible relation between the emission-line intensity distributions and the radio emission, we have overplotted the radio contours of the MERLIN radio image of \citet{mundell95} on the He{\sc\,i} flux map (top right panel of Fig.\,\ref{flux1}). While the radio intensity distribution is oriented horizontally in the Figure, along PA=75$\degr$, most intensity distributions are oriented instead along the [O{\sc\,iii}] bicone at P.A.=60/240$\degr$. Thus there seems not to be a strong association between the radio jet and the line emission.  On the other hand, both the radio and line emission are stronger to the SW than to the NE, and a closer inspection shows that the radio component C$_3$ (the radio knot just W of the nucleus) seems to align well with a region of enhanced emission 0$\farcs$4 W from the nucleus observed in the [Fe{\sc\,ii}]\,1.6440$\mu$m map.

\subsection{Line ratio maps}
\label{line_ratios}

We have used the intensity maps to construct the line ratio maps [S{\sc\,ii}]/[S{\sc\,iii}], [S{\sc\,iii}]/Pa$\beta$, [Fe{\sc\,ii}]\,1.257/Pa$\beta$, [Fe{\sc\,ii}]\,1.257/[P{\sc\,ii}], H$_2$\,2.122/Br$\gamma$, [Fe{\sc\,ii}]\,1.644/1.533, [Fe{\sc\,ii}]\,1.644/1.257 and  Br\,$\gamma$/Pa\,$\beta$, which are shown in Fig.\,\ref{ratio1}.  As pointed out above, at the nucleus (and  within $\approx$\,0$\farcs$2 from it), there may be some contribution from the broad-line component in Pa$\beta$ and Br$\gamma$, thus the nuclear ratios involving these two lines may be affected by this component. Many of these lines ratios are indicators of the gas excitation, thus the nuclear values should be used with caution.

The [S{\sc\,iii}]/Pa$\beta$ line-ratio map has values of $\approx\,4$ at the nucleus, increasing outwards to $\approx\,12$ to the SW and $\approx$\,8 to the NE. There is thus a systematic difference in the values of this line ratio between the two sides of the bicone: higher ratios in the near side (SW) and lower ratios in the far side (NE). The  [S{\sc\,ii}]/[S{\sc\,iii}] ratio map has the lowest values $<$0.1 at the nucleus, increasing outwards up to 0.2. 

We have overplotted the contours of the MERLIN radio image on the line ratio [Fe{\sc\,ii}]/Pa$\beta$ map 
in order to look for a possible relation between the [Fe{\sc\,ii}] excitation and the radio emission.There is indeed a relation: both the [Fe{\sc\,ii}]/Pa$\beta$
and [Fe{\sc\,ii}]/[P{\sc\,ii}] (Fig.\,\ref{ratio1}) line ratios increase outwards, reaching maximum values at $\approx$1\arcsec\,SW,  the location where the radio jet shows a ``flaring'' in its distribution, and to the opposite side at $\approx$1\arcsec\,NE, the location of a radio hotspot. The [Fe{\sc\,ii}]/Pa$\beta$ ratio increases from values $<1$ close to the nucleus up to $\approx$\,3 at the locations of the radio SW flare and NE hotspot, while [Fe{\sc\,ii}]/[P{\sc\,ii}] increases from values $\approx$\,4 to $\approx$\,8 at these same locations.  The line ratio [Fe{\sc\,ii}]\,1.644/1.533, which is sensitive to the emitting gas density, shows values $<4$ within the inner 0$\farcs$5, increasing to $\approx\,8-10$ in the outermost regions.

The H$_2$/Br\,$\gamma$ line ratio distribution is a result of the H$_2$ intensity distribution, which shows low or zero fluxes at and close to the nucleus and higher values perpendicular to the bicone. The consequence is an increase in the line ratio from values $\le$1 around the nucleus to $>$3 towards the borders of the H$_2$ emitting regions at $\approx$ 1\arcsec to the NW and SE of the nucleus. As in the case of Pa$\beta$, within 0\farcs3 from the nucleus, the line ratios may be affected by the broad component of the Br$\gamma$ line, and thus should not be used as indicator of the gas excitation. 

The line ratios [Fe{\sc\,ii}]\,1.644/1.257 and  Br\,$\gamma$/Pa\,$\beta$ are shown in the bottom panels of Fig.\,\ref{ratio1}. These line ratios can be, in principle, used for estimates of the reddening along the NLR, although the wavelength baseline is small, mainly for the [Fe{\sc\,ii}] ratio. The highest values are observed at the nucleus, for both ratios. In the [Fe{\sc\,ii}]\,1.644/1.257 ratio map there is a hint of a structure  resembling the  curved strands observed in the [O{\sc\,iii}] image (whose contours are overplotted on this map) in which the highest ratios seem to be observed in the regions of lowest [O{\sc\,iii}] fluxes. Data with better spatial resolution and higher signal-to-noise ratio would be necessary in order to confirm this result.

\section{Discussion}
\label{discussion}

\subsection{Ionized gas distribution}

The intensity distributions in the emission lines of the ionized gas (with the exception of  the [S\,{\sc viii}],  [S\,{\sc ix}], [Ca\,{\sc viii}] and [S\,{\sc vii}] coronal lines) resemble that of the optical  [O{\sc iii}] emission line, suggesting a similar origin, namely emission from ionized gas outflowing along the walls of a hollow bicone centered at the nucleus and oriented along PA=60/240$\degr$ \citep{hutchings99,das05,crenshaw00}. 
Nevertheless, as pointed out by \citet{kra08}, these intensity distributions do not clearly delineate a bicone, showing also extended emission perpendicular to the axis of the bicone in the vicinity of the nucleus. In other words, there is line emission beyond the presumed walls of the bicone close to the nucleus, indicating some escape of radiation in the perpendicular direction.

The fact that the light distributions are more extended and somewhat brighter to the SW than to the NE can be understood as due to the SW cone being directed toward us and  we are looking inside  the cone, where we are observing  the gas most exposed to the nuclear radiation field. To the NE we are observing the emission through both  the wall of the far  cone and the galactic plane (see discusssion below and  in Paper II).

\begin{figure}
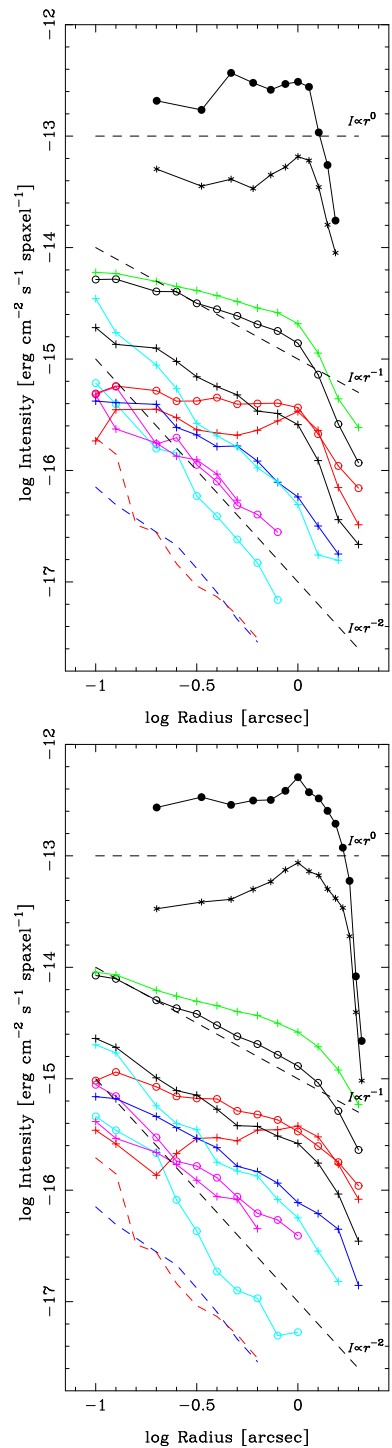

\centering
\includegraphics[angle=0,scale=0.43]{NGC4151_All_4_NE-1.ps}
\hspace{1cm}
\includegraphics[angle=0,scale=0.43]{NGC4151_All_4_SW-1.ps}
\caption{Radial emission-line profiles in cones of 40$^\circ$ opening
angle at PA = 60$^\circ$ ({\em top}) and 240$^\circ$ ({\em bottom})
along the bicone axis; [Fe{\sc\,ii}] 1.257 $\mu$m ({\em red
crosses}), [S{\sc\,ii}] 1.029/32/33 $\mu$m ({\em red open circles}),
H{\sc\,i} P$\beta$ 1.282 $\mu$m ({\em black crosses}), [S{\sc\,iii}]
0.9533 $\mu$m ({\em green crosses}), He{\sc\,i} 1.083 $\mu$m
({\em black open circles}), He{\sc\,ii} 1.013 $\mu$m ({\em dark blue
crosses}), [O{\sc\,ii}] 0.3727$\mu$m ({\em black asterisks}),
[O{\sc\,iii}] 0.5007$\mu$m ({\em black filled circles}), [Si{\sc\,vii}]
2.483 $\mu$m ({\em light blue crosses}), [Ca{\sc\,viii}] 2.322
$\mu$m ({\em light blue circles}), [S{\sc\,viii}] 0.991 $\mu$m ({\em
magenta crosses}), [S{\sc\,ix}] 1.2523 $\mu$m ({\em magenta open
circles}). Telluric star profiles in the J and K bands are shown in blue
and red dashed lines, respectively. Black dashed lines illustrate the 
slopes corresponding to different radial dependences for the intensity.}
\label{f:profiles}
\end{figure}

Radial profiles of the NLR emission have been obtained by
averaging the line flux within conical regions having an opening angle
of 40$^\circ$ centered on the nucleus and oriented along the bicone
axis at PA = 60$^\circ$ and 240$^\circ$. These radial profiles are
shown in Fig. \ref{f:profiles}. Profiles of [O\,{\sc ii}]\,3727\AA\ and [O\,{\sc iii}]\,5007\AA\ are also plotted from data presented by \citet{kra08}. All profiles decline steeply
beyond the edge of the bright inner NLR at radii between 1\arcsec\ and
2\arcsec. Among the near-IR profiles, the strongest features are those of 
[S\,{\sc\,iii}] 0.9533\,$\mu$m (green crosses) and He\,{\sc i} 1.083\,$\mu$m (black open circles).

Three distinct behaviours can be observed for the radial profiles 
in Fig\,\ref{f:profiles}. The first is observed for  [S\,{\sc\,iii}] 0.9533\,$\mu$m (green crosses) and 
the recombination lines of He\,{\sc i}\,1.083\,$\mu$m  (black open circles), Pa\,$\beta$ (black crosses) and 
He{\sc\,ii}\,1.013$\mu$m (dark blue crosses). 
Their intensity profiles decrease monotonically with distance from the nucleus,
showing a dependence on the radial distance of $\approx I\propto\,r^{-1}$, 
as illustrated by the black dashed lines in Fig.\,\ref{f:profiles}.
In order to understand the behavior of the lines, it would be necessary to construct models
to try to constrain the physical parameters which lead to line emission, what is beyond the scope of
the present study. A preliminary comparison of the radial profiles with models by \citet{gro04a} and
\citet{gro04b} suggest that  profiles such as the above are  reproduced by models including dusty 
radiation-pressure dominated clouds. Dust-free models for the NLR result in radial profiles for the
Pa\,$\beta$, He\,{\sc i} and He{\sc\,ii} emission lines which are steeper than those observed.

A distinct behaviour is observed in the  [Fe\,{\sc\,ii}] radial profile (red crosses in Fig.\,\ref{f:profiles}), 
which is enhanced at radii around $\sim$ 1\arcsec\ due to the
emission clumps that are apparent in Fig.\,\ref{flux2}.  This could be explained by excess gas-phase
abundance caused by shocks produced by the radio jet that has destroyed dust grains and released Fe.
Support for this interpretation is given by the correlation between by the  [Fe\,{\sc\,ii}] /[P\,{\sc\,ii}] 
ratio and the radio jet. In Paper II, we also find a spatial  correlation between the radio jet and 
the  [Fe\,{\sc\,ii}]  kinematics. The behavior of  [Fe\,{\sc\,ii}]  emission can also be partly due to the
fact that  [Fe\,{\sc\,ii}]  is produced in partially 
ionized regions beyond the main hydrogen ionization front in NLR clouds. Progressive 
absorption of FUV photons near the hydrogen ionization edge by absorbing clouds located 
between the central black hole and the NLR hardens the photoionizing spectrum. It is 
possible that the NLR has become optically thick to hydrogen-ionizing FUV radiation in 
its outer parts while it remains optically thin to X-rays from the AGN. In these 
circumstances, the outer NLR clouds could develop extensive partially ionized regions 
that emit relatively more [Fe\,{\sc\,ii}]  than P\,$\beta$. 

 [O\,{\sc\,ii}], [O\,{\sc\,iii}] and  [S{\sc\,ii}] (black asterisks, black filled circles and red circles, respectively)
 show a similar behaviour to that of [Fe\,{\sc\,ii}] , which can be approximately described  as $I\propto\,r^0$.

A third behaviour is observed in the coronal lines, which we now discuss. 

\subsection{Coronal gas distribution}

\begin{figure}
\includegraphics[scale=0.45]{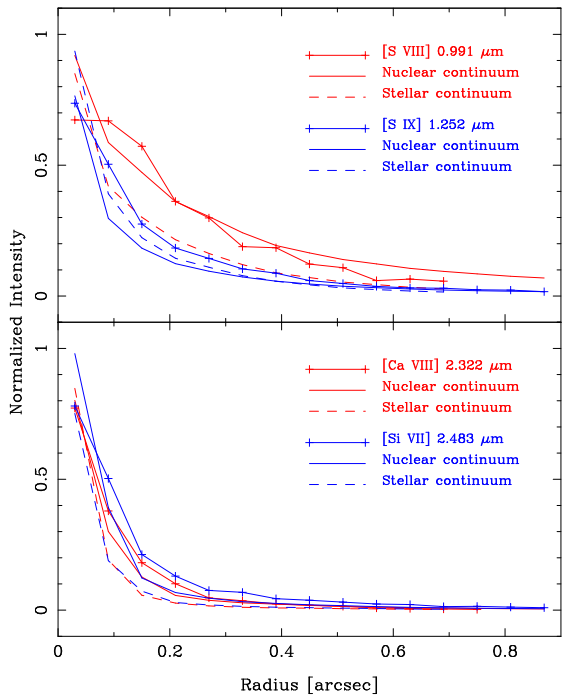}
\caption{Comparison of the spatial profiles of a star in the J and K-bands (dashed lines) with those of the galaxy nucleus in the J and K continuum and in the coronal lines  [S{\sc\,ix}], [S{\sc\,viii}], [Ca{\sc\,viii}] and [Si{\sc\,vii}].}
\label{psf}
\end{figure}

Fig. \ref{f:profiles}  shows that the coronal lines 
(light blue and magenta symbols) all decline more
steeply with radius than the recombination lines. In fact, they are
well-represented by an $I \propto R^{-2}$ intensity decline,
similar to that observed for the star radial profiles 
(shown as dashed blue and red lines in Fig.\,\ref{f:profiles}).
The [Ca{\sc\,viii}] feature (light blue circles in Fig. \ref{f:profiles})
declines even more steeply at small radii.

In order to verify if the coronal emission
is resolved and quantify its  extent, we have constructed azimuthal averages 
within circular radial annuli of the light distribution in each coronal line and have normalized the profile to unity at the peak. The resulting spatial profiles are shown in Fig.\,\ref{psf} together with those in the galaxy continuum and  telluric standard stars. In the top panel we compare the profiles of the  [S{\sc\,viii}]\,$\lambda$\,0.991\,$\mu$m and  [S{\sc\,ix}]\,$\lambda$\,1.252$\mu$m coronal lines with those in the continnum and telluric star in the Z and J bands, respectively. In the bottom panel, we compare the profiles of the [Ca{\sc\,viii}]\,$\lambda$2.322$\mu$m and [Si{\sc\,vii}]\,$\lambda$2.483$\mu$m coronal lines with those in the continnum and telluric star in the K and K$_{long}$ bands, respectively. 
As discussed in Sec.\,\ref{data}, if we assume that the PSF is given by the profile in the galaxy continuum, instead of the stellar profiles (except for the J-band, where we adopt the broader stellar profile), one concludes that none of the coronal lines is strongly resolved, except for  [Si{\sc\,vii}], which is somewhat extended along the bi-cone axis, what can be seen already in Fig.\,\ref{flux2}.  [Ca{\sc\,viii}] may be slightly 
more extended than the nuclear continuum, but higher spatial resolution and better 
sampled data, possibly with better signal-to-noise ratio, would be required to confirm 
this given that the NIFS spaxels have a size of 0$\farcs$1 in one direction.

The fact that the light distributions in  [S{\sc\,ix}]  and possibly [Ca{\sc\,viii}] coronal lines are extended, support an origin for these coronal lines in the inner part of the NLR or in the transition region between the BLR and the NLR, as suggested by previous authors  \citep[e.g.][]{ardila06}.
Ionization potentials for the parent species of each coronal line emitter are listed in Table
\ref{t:ip}. These generally support the formation of the observed
stratified coronal emission region through photoionization by the
central AGN, with higher ionization potential species located closer
to the ionizing source. With photon energies of 127.2\,eV being
required to create Ca{\sc\,viii}, the smaller extent of the [Ca{\sc\,viii}]
emission region when compared to that of the [Si{\sc\,vii}], 
which has higher ionization potential is unexpected. This fact can be attributed to depletion of
calcium onto dust grains, which can alter the gas-phase calcium
abundance by large factors \citep[e.g.][]{gro04a,gro04b}. 


\begin{table}
\caption{Ionization Potentials}
\centering
\begin{tabular}{l r}
\hline
~Line & IP (eV) \\
\hline
~[Fe{\sc\,ii}]   &   7.9 \\
~[S\,{\sc ii}]    &  10.4 \\
~[P{\sc\,ii}]     &  10.5 \\
~[S{\sc\,iii}]   &  23.3 \\
~[Ca{\sc\,viii}] & 127.2 \\
~[Si{\sc\,vii}] & 205.3 \\
~[S{\sc\,viii}]  & 280.9 \\
~[S{\sc\,ix}]    & 328.8 \\
\hline                                                                                                
\end{tabular}
\label{t:ip}                                                                                        
\end{table}




\subsection{Molecular gas distribution}


The H$_2$ intensity distribution is totally different from those of
the other emission lines. It avoids the bicone (bottom left panel of
Fig.\,\ref{flux2}) and is more extended than other emission lines along the minor
axis of the galaxy, which coincides approximately with the orientation
of the bar. The H$_2$ emission covers a region $\approx$ 20-60 pc in
radial extent from the nucleus. One possible explanation for this
morphology is that the nuclear molecular gas is located predominantly
in the plane of the galaxy, but is dissociated by the AGN radiation
field within the bicone. It is known that the bicone intercepts part
of the galaxy plane \footnote{See Paper II and  \citet[e.g.][] {das05}
for a comprehensive discussion of the kinematics and geometry of the
bicone and the galaxy.}. Perpendicular to the bicone, the H$_2$ must
be shielded from the strongest dissociating radiation, probably by a
dusty torus and/or by the walls of the bi-conical outflow \citep{kra08}. 

A previous H$_2$$\lambda$2.1218$\mu$m image of the nuclear region of NGC\,4151 was otained by \citet{fernandez99} who have found a similar intensity distribution to ours, in the form of a partial ring surrounding the nucleus, which led them to propose that the H$_2$ emission originated in the outer part of a molecular torus. They suggested that  the H$_2$ emitting gas could  be rotating in a plane perpendicular to the radio axis. We note, however, that the orientation of the partial ring shown in \citet{fernandez99} is rotated by 90$\degr$ and is mirrored relative to our H$_2$ intensity distribution. Otherwise, the intensity distributions are consistent with each other considering the difference in spatial resolution (0$\farcs$6 in \citet{fernandez99} and 0$\farcs$11 here). As our datacube gives intensity distributions for the ionized gas in agreement with that for the well known bicone, and the H$_2$ intensity distribution is obtained from the same datacube, we conclude that the orientation of the H$_2$ ring is mistaken in \citet{fernandez99}. Our observations also do not support the suggestion by these authors  that the H$_2$ gas is rotating in a plane perpendicular to the radio jet. In Paper II we present  the H$_2$ kinematics which shows very little rotation  consistent with a gas distribution in the galaxy plane along the minor axis. Nevertheless, our observations are consistent with the interpretation that  the H$_2$ emitting gas  may be tracing  the gas reservoir which feeds the SMBH. Results supporting this idea are the inflows  measured  by \citet{mundell99a}  in radio observations of H\,{\sc i} along the  large scale bar (at PA=130$\degr$). These inflows direct gas towards  the inner part of the bar in the nuclear region and their orientations are approximately perpendicular to the arc-shaped structures delineated by the H$_2$ light distribution (see Fig.\,\ref{flux2}). Thus  one possibility is that the radio observations are tracing H\,{\sc\,i} inflows  towards  the nuclear region leading to  the  build up of the molecular gas reservoir which we observe in the H$_2$ intensity distribution.

\subsection{Extinction}

Previous studies report low reddening towards the NLR of NGC\,4151.
\citet{crenshaw05} claim only low Galactic
extinction of $E(B-V) = 0.02\pm0.04$ mag in
STIS spectra centered on the ``nuclear emission-line knot''. 
\citet{kra00} report that at other locations in the NLR the optical reddening of the
emission lines ranges from $E(B-V)\approx 0.0$ to 0.4.

Other studies \citep[e.g.][]{alex99} report reddening estimates 
which range from almost negligible
$E(B-V) = 0.04-0.05)$ \citep[e.g.][]{kriss95,penston81} to the considerable $E(B-V)
= 0.13$\,\citep{malkan83}.  From a variety of methods (and data), including
near-IR emission lines, \citet{ward87} get $E(B-V) \sim 0.23$ mag, 
while \citet{rl81} obtain $0.5 < A_V< 0.8$. 

\citet{mundell95} measure H{\sc\,i} absorption
to the radio component C$_4$ of $3.9 \times 10^{21}$ cm$^{-2}$. From \citet{bsd78}
$<N(HI+H_2)/E(B-V)> = 5.8 \times 10^{21}$ cm$^{-2}$
mag$^{-1}$, so this H{\sc\,i} column corresponds to $E(B-V) = 0.7$
($A_V \sim 2.1$ mag) to the nuclear radio source, if there is no
intervening H$_2$. 


We can use the line ratio map [Fe{\sc\,ii}] $\lambda$1.2570/$\lambda$1.6440 to estimate the
reddening to the NLR of NGC\,4151. 
These lines arise from the same upper level so the
intrinsic value of the line ratio should be 1.36, according to the
transition probabilities of \citet{ns88}  and as
confirmed from observations by  \citet{baut98}. Smaller
line ratios indicate the presence of reddening and can be used to
estimate its value through the relation:


\begin{equation}
E(B-V)=8.14\times log\left(\frac{1.36}{\frac{F_{1.2570}}{F_{1.6440}}}\right)
\end{equation}

\noindent obtained using the reddening law of \citet{ccm89}.

Typical values of the reddening uncertainty can be obtained from the
data in Table\,\ref{fluxes}. The measured flux ratios are $1.21\pm0.06$
and $1.25\pm0.12$ for the SW  [Fe{\sc\,ii}] peak and Position B, respectively.
These ratios equate to $E(B-V) = 0.41\pm0.14$ and $0.30\pm0.28$,
from which we estimate random uncertainties in
$E(B-V)$ derived by this method to be $\approx \pm0.2$ mag along the radio axis.
For the nucleus and other regions where the  [Fe{\sc\,ii}] emission is weaker,
the uncertainties are twice as large.
 
\begin{figure}
\includegraphics[scale=0.42]{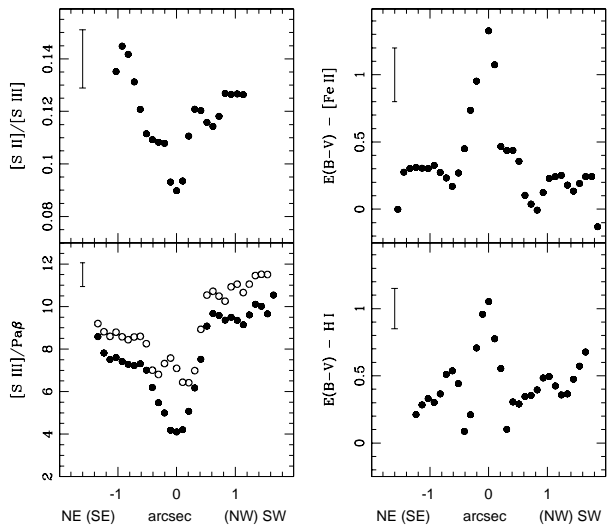}
\caption{Left: spatial line ratio profiles along the bicone axis (PA$=60\degr$) obtained from smoothed 1D profiles extracted within a pseudo slit of width 0$\farcs$3. Filled circles show values not corrected for reddening, while the open circles show values corrected for reddening with E(B-V) obtained from the average of the two E(B-V) values from the right panels for each location. Right: E(B-V) values obtained from the line ratio profiles of [Fe{\sc\,ii}]$\lambda$1.257/1.644 (top) and  Pa\,$\beta$/Br\,$\gamma$ (bottom).}
\label{ratio2}
\end{figure}

In order to investigate further the reddening variation across the NLR, we
have extracted smoothed 1D profiles from the [Fe{\sc\,ii}]$\lambda$1.644 and 1.257$\mu$m 
intensity maps along a pseudo slit of width 0$\farcs$3, oriented along the bicone. The fluxes were obtained at each 0$\farcs$1 along the pseudo slit as the average of  the fluxes of the pixels included within the 0$\farcs$3 width of the slit. The 1D profiles were then smoothed further by replacing the flux at each position by the average of its flux and those of the two adjacent positions along the slit. We have then constructed the ratio between the two 1D profiles along the bicone and obtained the $E(B-V)$ from the expression above. The result is shown in the right upper panel of Fig.\,\ref{ratio2}, showing a maximum $E(B-V)=1.4$ at the nucleus, which  decreases abruptly to values in the range $0.3<E(B-V)<0.6$ beyond 0$\farcs$2 from the nucleus, with an average value of $E(B-V)\approx\,0.45$.

In order to check the above result, we have next used the Pa\,$\beta$/Br\,$\gamma$ ratio to estimate the reddening along the same pseudo slit, using the relation:

\begin{equation}
E(B-V) = 4.74 \times log\left(\frac{5.88}{\frac{F_{P\beta}}{F_{Br\gamma}}}\right)
\end{equation}

\noindent where we have used again the reddening law of \citet{ccm89} and adopted the intrinsic $F_{P\beta}/F_{Br\gamma}$ ratio  of 5.88 corresponding to case B recombination \citep{ost89}. Due to the somewhat longer wavelength baseline, the uncertainties in the resulting $E(B-V)$, of $\pm$\,0.15, are smaller than that obtained using the [Fe{\sc\,ii}] ratio. The results are shown in the bottom right panel of Fig.\,\ref{ratio2}. Within the uncertainties, the $E(B-V)$ values and spatial variation are similar to those obtained using the [Fe{\sc\,ii}] ratio: the highest values are observed at the nucleus, falling abruptly outwards to an average value around $E(B-V)\approx$\,0.5.

We note that the nuclear E(B-V) values are in agreement with the one obtained by \citet{mundell95} for the nuclear source, while E(B-V)$\approx$0.5 is consistent with the highest E(B-V) values obtained by \citet{kra00}, considering also the fact that the optical depth reached in the near-IR is larger than those obtained by optical observations. Finally, we pointed out at the end of Section\,\ref{line_ratios} that there may be a relation between the [Fe{\sc\,ii}]$\lambda$1.257/1.644 map and the [O{\sc\,iii}] intensity distribution. In terms of reddening distribution, this relation would imply that the regions with largest reddening would coincide with the gaps in the intensity distribution, suggesting that the appearance of the [O{\sc\,iii}] intensity distribution could be, at least in part, due to reddening. As E(B-V)$\approx$0.5 has been obtained from smoothed profiles, this value is consistent with higher extinction in narrow structures not resolved by these smoothed profiles. For example, if the average E(B-V)$\approx$0.5 is due to the smoothing of narrow strands having  E(B-V)=1, with E(B-V)=0 between the strands, the [O{\sc\,iii}] flux will be reduced by a factor of $\approx$25  where E(B-V)=1, what could result in the stranded appearance observed in the  [O{\sc\,iii}]  intensity distribution. Data with better spatial resolution and higher signal-to-noise ratio would be necessary in order to test this possibility.




\subsection{Gas excitation}

The line ratio [S{\sc\,iii}]/Pa$\beta$ (top right panel of Fig. \ref{ratio1}) can be used as a tracer of the excitation of the NLR gas. We have obtained typical values for this line ratio in active and starburst galaxies from the works of  \citet{sb95} and \citet{riffel06a}. In the case of \citet{sb95}, we have used their observed [S{\sc\,iii}]/H$\alpha$ ratio values (not corrected for reddening) to estimate the [S{\sc\,iii}]/Pa$\beta$ ratios (assuming a typical reddening of E(B-V)=0.5), while in the case of \citet{riffel06a} we have used their measured values for  [S{\sc\,iii}] and Pa$\beta$ fluxes, also not corrected for reddening. In order to avoid the contribution of the broad components to the emission lines, we have collected only the values for Seyfert 2 galaxies as typical of the NLR, as well as those for starburst galaxies, to use as a comparison. The range of line ratios obtained for a total sample of 14 Seyfert 2 galaxies is $6 \le$ [S{\sc\,iii}] $\le 14$, with most values clustering around  [S\,{\sc iii}]$\approx$8.5, while for a sample of 10 starburst galaxies the observed range is $1 \le$ [S\,{\sc iii}] $\le 3.5.$. These values can be compared to those in Fig.\ref{ratio1}.  In order to make such a comparison easier, we present, in the bottom panel of Fig.\ref{ratio2},  line ratio profiles along the bicone axis  obtained from smoothed 1D profiles extracted along a pseudo slit of width $0\farcs$3, as described in the previous section. 

In the 1D profiles of fig.\,\ref{ratio2}, the observed [S{\sc\,iii}]/Pa$\beta$ line ratios (filled circles) beyond 0$\farcs$5 from the nucleus are in the range 7--10, thus similar to the values for NLR of Seyfert 2 galaxies, with the values to the NE being 20--30\% smaller than those to the SW. Within $0\farcs$5 from the nucleus the line ratio decreases down to half the values beyond this region. This decrease is partially due to reddening, as can be observed in the reddening corrected line ratio profile shown as open circles in the bottom left panel of Fig.\,\ref{ratio2}. The corrected values were obtained by using again the reddening law of \citet{ccm89} and the E(B-V)  for each location as the average of the two values shown in the right panels of Fig.\,\ref{ratio2}. The values of the line ratio within 0$\farcs$5 from the nucleus increase from 4--6 to 7--8, now within the typical range observed for AGN. Nevertheless, the correction for reddening does not eliminate completely the decrease observed at the nucleus, and another effect may be present, possibly a residual contribution of the broad component of H$\beta$.
In principle, the 20--30\% smaller [S{\sc\,iii}]/Pa$\beta$ values to the NE relative to the SW could also be  due to an excess reddening of E(B-V)$\approx$\,0.5. Nevertheless, the E(B-V) profiles in the right panels of Fig.\,\ref{ratio2} do not show a systematic difference in reddening between the SW and NE of this order, and we thus attribute the lower ratios to  NE to lower excitation. 

The [S{\sc\,ii}]/[S{\sc\,iii}] ratio (top left panel of Fig. \ref{ratio1}) can also be used as a tracer of the excitation, although this ratio is noisier than the [S{\sc\,iii}]/Pa$\beta$ ratio due to the fact that the [S{\sc\,ii}] flux map is a collection of the fluxes in four faint emission lines. We have also constructed a 1D line-ratio spatial profile for [S{\sc\,ii}]/[S{\sc\,iii}], which is shown in the top left panel of Fig. \ref{ratio2}. Beyond 0$\farcs$5 from the nucleus, this ratio is higher to the NE, which indicates lower excitation relative to  the SW side of the bicone, in agreement with the behavior of the [S{\sc\,iii}]/Pa$\beta$ ratio. Within 0$\farcs$5 from the nucleus,  the  [S{\sc\,ii}]/[S{\sc\,iii}] ratio decreases, what cannot be attributed to reddening, as  an excess reddening should increase the value of this line ratio and not decrease. Although this decrease could be due to a higher excitation in this region, this is not consistent with the behavior of the [S{\sc\,iii}]/Pa$\beta$ line ratio. Our favored explanation is as follows. In the next section we use the observed [Fe{\sc\,ii}]  line ratios to derive the electronic density along the NLR. Within $\approx\,0\farcs$5 from the nucleus, the average density (see Fig.\,\ref{fe_ratio} is higher than the critical density of  [S{\sc\,ii}] (2.5$\times\,10^4$\,cm$^{-3}$), resulting in a decrease in the intensity of the  [S\,{\sc ii}] lines close to the nucleus.

How can one understand the lower excitation to the NE as compared to the SW side of the cone? According to previous models \citep[e.g.][]{hut98,crenshaw00,das05},  the gas in the NLR is outflowing along  a hollow cone, with  the SW side tilted toward us. Our line of sight is almost along, but outside, the near edge of the SW cone. As the cone seems to have a geometrical cutoff at $\approx$100\,pc from the nucleus, we are looking at the inner wall of the cone to the SW, while to the NE we are looking at the outer wall of the cone, which is  at least partly behind the plane of the galaxy.  We see higher excitation when we look at the inner wall of the cone and somewhat lower excitation when we look at the outer wall of the cone.  As we do not see any reddening difference between the two sides of the cone, the only explanation seems to be that the ionizing radiation is attenuated more to the NE than to the SW.

Recently, \citet{kra08} have mapped the ionization in the NLR of NGC\,4151 using the line ratio map [O{\sc\,iii}]]/[{O\sc\,ii}] obtained from the ratio of  narrow-band HST images. They compare the observed emission-line ratios along the NLR with photoionization models and point out that, while the highest line ratios are observed along the bicone, there is also emission from regions outside the cone, with smaller line ratios, indicating lower excitation. They attribute this lower excitation to a weaker ionizing flux reaching these regions due to  the presence of a ``low-ionization absorber'' which filters the radiation. This absorber seems to be ionized gas, leading \citet{kra08} to propose that  the low-ionization absorbers are dense knots of gas swept up by an accretion disk wind. 

Our excitation maps do not seem to show a decrease of the excitation perpendicular to the bicone as observed by \citet{kra08}.  Instead, we find lower excitation to the NE than to the SW, as discussed above. Maybe the difference between our results and those of  \citet{kra08} can be attributed to the different optical depths probed by optical and near-IR observations. The difference in excitation between the NE and SW sides of the bicone could then be due to the presence of more absorbers between the nuclear ionizing source and the emitting gas we see to the NE than to the SW.


\subsection{Physical conditions of the [Fe{\sc\,ii}] emitting region}
\label{Fe}

\begin{figure}
\includegraphics[scale=0.34, angle=-90]{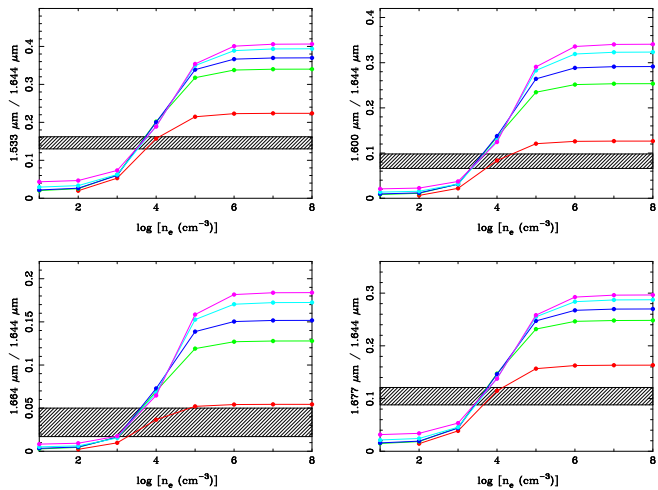}
\caption{Model [Fe{\sc\,ii}] emission-line ratios versus electron density. Each sequence of models correspond to a different temperature -- from bottom to top -- 1, 3, 5, 10 and 20\,$\times$\,10$^3$K. The shaded regions correspond to the measured line ratio and their uncertainties within a region of aperture 0$\farcs3\times0\farcs$3 centered in the [Fe{\sc\,ii}] emission peak at 0$\farcs$9\,SW from the nucleus.}
\label{Ne}
\end{figure}

There are many [Fe{\sc\,ii}] emission lines in our near-IR spectra of the  NLR of NGC\,4151, specially to the SW, where the [Fe{\sc\,ii}] emission is strongest. The  near-infrared [Fe{\sc\,ii}] emission lines are due to forbidden transitions between low energy levels, and their intensities are thus principally dependent on the electron density.  We were able to measure four  [Fe{\sc\,ii}]  line ratios in the  region of strongest [Fe{\sc\,ii}] emission (at $\approx$\,0$\farcs$9\,SW of the nucleus), from the fluxes listed in Table \ref{fluxes}, which can be used as density indicators. We have performed a 16-level atom calculation for [Fe{\sc\,ii}] and have obtained line intensities in order to compare with the observed ones in NGC\,4151 and derive the electron density.  Fig.\,\ref{Ne} shows the model results for four emission-line ratios as a function of the electron density for different values of the electronic temperature, along with the measured ratios and uncertainties. The line ratios indicate a density in the region of the strongest [Fe{\sc\,ii}] emission of $\approx$\,4000\,cm$^{-3}$. 

\begin{figure}
\includegraphics[scale=0.7]{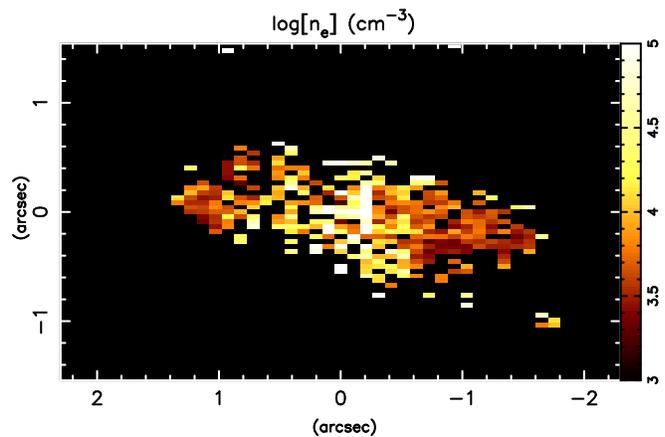}
\caption{Distribution of electron density n$_e$ obtained from the [Fe{\sc\,ii}] 1.533/1.644 emission-line ratio, for an electron temperature in the range 5000K$<T_e<$20000K.}
\label{fe_ratio}
\end{figure}

Most of the [Fe{\sc\,ii}] emission-line ratios that are density indicators could be measured only at the location of maximum [Fe{\sc\,ii}] emission. The exception is  [Fe{\sc\,ii}]1.533/1.644, which could be measured at several other locations allowing the construction of  an electron density map, shown in Fig.\ref{fe_ratio}. The uncertainty in the line flux of the 1.533$\mu$m line is quite large at the nucleus (see Table \ref{fluxes}), but  decreases outwards. It can be concluded from Fig.\,\ref{fe_ratio} that the electronic density in the  [Fe{\sc\,ii}] region decreases from the highest values at the nucleus  ($\ge\,10^5$\,cm$^{-3}$),  to $\approx$\,10$^4$\,cm$^{-3}$ within the inner $\approx$0$\farcs$5 from the nucleus. Outside this radius the density reaches the value of $\approx$4000\,cm$^{-3}$ obtained above.

Deriving the electron temperature in the [Fe{\sc\,ii}] emitting region is more difficult, as most observable emission lines arise from a limited range of upper level energies. Following  \citet{mouri00} and \citet{paper_1275}, we looked for emission lines corresponding to transitions from the a$^4$P term. There is a transition between the terms $^4$P$_{4/3}$ and a$^4$D$_{7/2}$ corresponding to an emission line at 1.7489$\mu$m in the H-band. Although we have apparently detected this line, its uncertainty is too large to put a useful contraint on the electron temperature. Another possibility suggested by \citet{thompson95} 
is to use the [Fe{\sc\,ii}]\,0.8619$\mu$m line strength, corresponding to the transition a$^4$P$_{5/2}$--a$^4$F$_{9/2}$ as an indicator of the electron temperature. Unfortunately, our observed spectral range does not cover this emission line and we had to look for previous observations available in the literature in which this line was measured. \citet{ost90} have identified the [Fe{\sc\,ii}]\,0.8619$\mu$m emission line in their spectrum of NGC\,4151, reporting a flux ratio [Fe{\sc\,ii}]\,0.8619/[S{\sc\,iii}]\,0.9533=0.042. We have used  this ratio and our measurement of the [S{\sc\,iii}]\,0.9533 line flux at 0$\farcs$9\,SW from the nucleus in order to estimate the flux of  the  [Fe{\sc\,ii}]\,0.8619 emission line at this location, under the assumption that the [Fe{\sc\,ii}]\,0.8619/[S{\sc\,iii}]\,0.9533 ratio is the same as that obtained by \citet{ost90}, in spite of their much larger aperture. Using the fluxes from Table \ref{fluxes}, we obtain [Fe{\sc\,ii}]\,0.8619=2.09$\times10^{-15}$\,ergs\,cm$^{-2}$\,s$^{-1}$, and thus a ratio [Fe{\sc\,ii}]\,1.2570/0.8619=3.6. If we consider a reddening value  E(B-V)$\approx$0.5, the corrected log([Fe{\sc\,ii}]\,1.2570/0.8619)=0.46$\pm$0.11 and the resulting temperature will be 15,000\,$\pm$\,5000\,K (considering the uncertainties in the line ratio combined with those in the reddening correction).  

\begin{figure}
\includegraphics[scale=0.36,angle=-90]{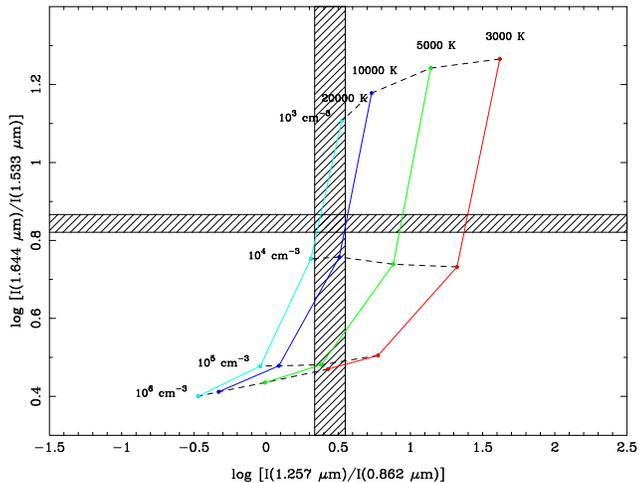}
\caption{Model [Fe{\sc\,ii}]   line ratio 1.644$\mu$m/1.533$\mu$m (which depends mostly on the gas density)  versus the line ratio 1.257$\mu$m/0.862$\mu$m (which depends mostly on the temperature) as a function of density and temperature. The shaded region corresponds to the measured line ratios and their uncertainties within a region of aperture 0$\farcs3\times0\farcs$3 centered in the [Fe{\sc\,ii}] emission peak at 0$\farcs$9\,SW from the nucleus.}
\label{fe_ratio_ratio}
\end{figure}

The origin of [Fe{\sc\,ii}] emission in galaxies has been the subject of many studies, since the finding by \citet{forbes93} and \citet{blietz94} of a correlation between the [Fe{\sc\,ii}] and radio emission. This suggests that shock excitation by radio jets is a likely mechanism for production of the [Fe{\sc\,ii}] emission \citep{dopita95,dopita96}. However, subsequent works have argued that the dominant excitation mechanism is photoionization, with shock excitation accounting for only $\approx$20\% of the [Fe{\sc\,ii}] excitation in AGN \citep{simpson96}. 

An important tracer of the origin of the [Fe{\sc\,ii}] emission is the line ratio [Fe{\sc\,ii}]/Pa$\beta$, whose value ranges from $<$0.6 for starbursts to $>2$ for supernova remnants, for which shocks are the dominant mechanism \citep{ardila04}. Active galaxies have values for this ratio between 0.6 and 2, suggesting that for ratios close to 0.6 photoionization is the dominant mechanism, while for ratios close to 2, shock excitation dominates \citep{sb99,ardila04}. [Fe{\sc\,ii}]/Pa$\beta$ ratios $\ge$2 have been found by \citet{riffel06b} in a near-IR IFU study the NLR of the Seyfert 2 galaxy ESO\,428--G14 in regions showing a close association with emission knots of a radio image, supporting an origin for the [Fe{\sc\,ii}] emission in shocks produced by the radio jet.

In NGC\,4151, [Fe{\sc\,ii}]/Pa$\beta\le$1 within the inner 0$\farcs$6, increasing to values of $\approx$2 outwards (see Fig.\ref{ratio1}), suggesting that photoionization dominates in the inner region while shocks dominate in the outer region. Indeed, inspection of the [Fe{\sc\,ii}]/Pa\,$\beta$ ratio map in Fig.\,\ref{ratio1} shows a relation with the radio structure: to the SW, the increase in the [Fe{\sc\,ii}]/Pa$\beta$ ratio does coincide with a flaring in the radio contours, which may have been produced by a shock between the jet and ambient gas, while to the NE, a similar increase is observed at the location of a radio knot. 

Another powerful indicator  of the origin of the [Fe{\sc\,ii}] emission is the [Fe{\sc\,ii}]\,1.2570/[P{\sc\,ii}] line ratio, as pointed out by \citet{oliva01}. This line ratio is seldom used because the [{P{\sc\,ii}] emission line is usually too faint. The quality of our data has allowed the measurement of extended emission in this line. Besides having similar wavelengths, the [Fe{\sc\,ii}]\,1.2570 and [P{\sc\,ii}] lines have similar excitation temperatures, and their parent ions have similar ionization potentials and radiative recombination coefficients. \citet{oliva01} have shown that, for a solar abundance, the above ratio is $>20$, as observed in supernova remnants. But, in many astronomical objects this ratio is much smaller, because Fe is locked into grains. For example, in Orion, the above ratio is $\approx$2. In order to destroy the dust grains and release the Fe, fast shocks are necessary. Larger ratios  than 2 indicate that shocks have passed through the gas  destroying the dust grains, releasing the Fe and thus enhancing its observed abundance. \citet{oliva01} have reported the first observation of this line ratio in an extragalactic source, and here we report the first 2D map of this ratio in an extragalactic source.
In NGC\,4151, [Fe{\sc\,ii}]/[P{\sc\,ii}] varies from 2 to 6, with the highest values observed in patches which do seem to be spatially correlated with the radio structure, as shown in Fig.\,\ref{ratio1}: the highest ratios are observed at the location of a radio knot to the NE and at $\approx\,0\farcs$8 SW from the nucleus, where the flaring of the radio contours is observed. These locations also coincide with those where the  ratio [Fe{\sc\,ii}]/Pa$\beta$ shows the highest values, supporting an increased contribution of shocks to the excitation of [Fe{\sc\,ii}] at these locations. Nevertheless, the [Fe{\sc\,ii}]/[P{\sc\,ii}] line ratio is never as high as $\approx$20, which suggests  that the excitation mechanism is not only shocks, but includes photoionization.

\subsection{Physical conditions of the H$_2$ emitting region}

As discussed by \citet{riffel06b}, the H$_2$ emission lines can be excited by two processes: (1) fluorescence by soft-UV photons \citep{black87} -- present both in star-forming regions and around active nuclei; or (2) thermal processes, produced either by  X-ray \citep{maloney96} or shock heating \citep{hollenbach89}.

An emission-line ratio commonly used to investigate the origin of the H$_2$ excitation is H$_2$\,2.1218/Br\,$\gamma$. In starburst galaxies, where the main heating agent is UV radiation, H$_2$/Br$\gamma<0.6$ \citep{ardila04}, while for Seyferts this ratio is larger because of additional thermal excitation by shocks or  X-rays from the AGN. As observed in  Fig.\,\ref{ratio1}, this line ratio is $<$1 around the nucleus and along the axis of the bicone, but increases to values $>2$ and up to 4 along the minor axis of the galaxy (and approximately perpendicular to the bicone axis). By comparing Fig.\ref{ratio1} with Fig.\ref{flux2} which shows that the H$_2$  intensity distribution avoids the region along the bicone axes, we propose that the low H$_2$/Br$\gamma$ ratio in these regions is due to the destruction of the H$_2$ molecule by the strong radiation and particle fluxes, while the high values at the perpendicular direction could be due to fluorescence or to excitation by nuclear X-rays escaping in that direction or by shocks. In the case of NGC\,4151, shocks in the region where H$_2$ emission is observed, perpendicular to the radio jet, can be produced in gas flowing into the nuclear region  or associated with star formation in the disk. It could be even possible that star formation is occurring in in-flowing gas. 

The H$_2\,\lambda2.2477/\lambda2.1218$ line ratio can be used to distinguish between thermal ($\sim$0.1-0.2) and fluorescent ($\sim$0.55) excitation \citep{mouri94,reunanen02,ardila04}. The first line could be measured in a few locations along the NLR. The resulting line ratio 
does not vary much along the NLR, and the average  value is 0.13$\pm$0.02, supporting a thermal excitation for the H$_2$.

We can investigate further the excitation mechanism of H$_2$ by using all its emission-line fluxes in the $K$ band to calculate  its excitation temperature $T_{exc}$ and the ratio of the {\it ortho}  to {\it para} emission lines. As the H$_2$ emission-line ratios seem not to show spatial variation, in order to improve the signal-to-noise ratio, we have measured the H$_2$ fluxes  within a 0$\farcs$5 diameter aperture centered on the SE peak of the H$_2$ intensity distribution. Following \citet{wilman05}, we have investigated the relation:

\begin{equation}
log\left(\frac{F_i \lambda_i}{A_i g_i}\right)=constant - \frac{T_i}{T_{exc}}
\end{equation}

\noindent where $F_i$ is the flux of the $ith$ H$_2$ line, $\lambda_i$ is its wavelength, $A_i$ is the spontaneous emission coefficient, $g_i$ is the statistical weight of the upper level of the transition and $T_i$ is the energy of the level expressed as a temperature. This relation is valid for  thermal excitation, under the assumption of an $ortho:para$  abundance ratio of 3:1. $T_{exc}$ (the reciprocal of the slope) will be the kinetic temperature if the H$_2$ is in thermal equilibrium. 

The above relation is plotted in  Fig.\,\ref{T_exc} where we have included the fluxes of most H$_2$ emission lines, excluding only the weakest ones (1-0 Q(6), 1-0 Q(7) and 2-1 S(1)).  The result is a tight straight line with $T_{exc}=$2155\,K. The fact that all emission lines are on the relation confirms that the H$_2$ is in thermal equilibrium at $T_{exc}$, and that the rotational and vibrational temperatures are the same, ruling out a significant contribution from fluorescence. The fact that the $ortho$ lines (1-0 S(1), 2-1 S(1), 1-0 Q(1), etc) give the same result as the $para$ lines (1-0 S(0), Q(2), Q(4), etc) confirms that the $ortho$ to $para$ ratio is $\sim 3$ as assumed. This is also what is  expected for thermal equilibrium.

We have thus concluded that the H$_2$ emission-line ratios are consistent with thermal excitation, which excludes UV fluorescence from the nucleus as a mechanism to excite the H$_2$ emission. But there are two other possibilities: heating by nuclear X-rays  or by shocks in the gas flowing to the nucleus.

In order to test if X-rays from the AGN can account for the observed H$_2$ line fluxes, we have used the models of \citet{maloney96} to estimate the H$_2$ flux emitted by a gas cloud illuminated by a source of hard X-rays with luminosity $L_X$. The calculations are performed as described in \citet{zuther07} and \citet{riffel08}. Using a X-ray flux of $4.51\times\,10^{-11}$\,erg\,cm$^{-2}$\,s$^{-1}$ and an absorption column density of  $7.5\times\,10^{22}$\,cm$^2$ for the nuclear source \citep{cappi06}, we obtain a X-ray flux at a typical distance from the nucleus of 33\,pc (0$\farcs5$ from the nucleus, where H$_2$ emission is observed) of 7.3\,erg\,cm$^{-2}$\,s$^{-1}$ and an effective ionization parameter of $\xi_{\rm eff}=0.015$. Using Fig.\,6a from \citet{maloney96}, we can then estimate the resulting H$_2$ flux at the Earth for an aperture corresponding to our pixel of $0\farcs1\times0\farcs04$ -- which gives a solid angle of $9.6\times\,10^{-14}$\,sr -- as $7\times\,10^{-15}$\,erg\,cm$^{-2}$\,s$^{-1}$. Inspection of Fig.\ref{flux2} shows that the highest values observed are $\approx 10^{-16}$\,erg\,cm$^{-2}$\,s$^{-1}$ and we conclude that excitation by X-rays emitted by the AGN can account for most of the observed H$_2$ flux if a similar X-ray flux to that escaping in our direction also reaches the region of H$_2$ emission.

\citet{prestwich92} analysed near-infrared spectra of NGC\,4151
obtained with $>$ 5.4\arcsec\ diameter apertures over a ten year period from
1979 to 1989 and found that the H$_2$ line emission was stable over this
interval while the (broad) H{\sc\,i} Br$\gamma$ emission varied
significantly. This could argue against nuclear X-ray heating. However, the
light crossing time of the H$_2$-emitting region with its $\sim$ 1\arcsec\
radial extent (Fig.\,\ref{flux2}) is $\sim$ 200 yr, so nuclear variations on decade
timescales will be spatially diluted in large aperture measurements.
Observations of individual H$_2$ clumps at higher angular resolution than
possible with NIFS would be required to provide tighter constraints since
the light crossing time of a 0.1\arcsec diameter clumps is still $\sim$ 20\,yr.

What then is powering the H$_2$ emission? Our observations are consistent with two mechanisms: X-rays from the active nucleus
and/or shocks in the inner galactic disk, possibly produced by the  accretion flow along the large scale bar \citep{mundell95}, which may be the origin of the H$_2$ gas accumulated in the inner region.

\begin{figure}
\includegraphics[scale=0.32,angle=-90]{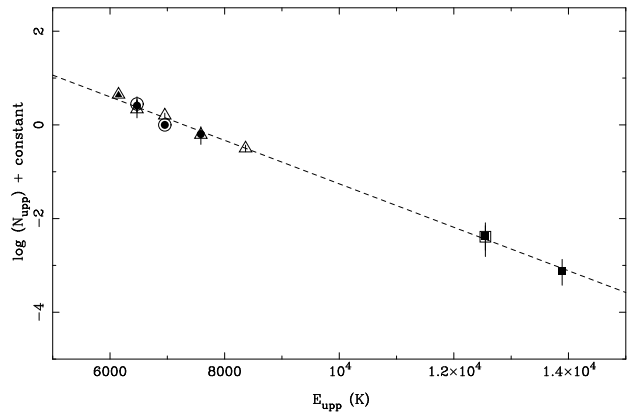}
\caption{Relation between $N_{upp}=\frac{F_i \lambda_i}{A_i g_i}$ and $E_{upp}=T_i$ for the H$_2$ emission lines for thermal excitation at temperature $T_{exc}=$2155\,K. Transitions plotted are from (left to right) 1-0 Q(1), 1-0 S(0), 1-0 Q(2), 1-0 S(1), 1-0 Q(3), 1-0 S(2), 1-0 Q(4), 1-0 Q(5), 2-1 S(1) and 2-1 S(3).}
\label{T_exc}
\end{figure}


\subsection{Mass of ionized and molecular gas}

The mass of ionized hydrogen can be estimated as $M_{\rm HII}=m_p\,N_e\,V_{\rm HII}$, where $N_e$ is the electron density and $V_{\rm HII}$ is the volume of the emitting region. 

The product  $N_e^2\,V_{\rm HII}$ can be obtained from the expression for the Br$\gamma$ flux obtained using the H{\sc\,i} emission coefficients listed in \citet{ost89}:

\begin{equation}
F_{\rm Br\gamma}=2.7\times10^{-28}\frac{N_e^2 V_{\rm HII}}{D^2}~~~~~~~~~{\rm erg\, cm^{-2}\, s^{-1}}
\end{equation}

\noindent
where $D$ is the distance to the galaxy in cm, units of density are cm$^{-3}$ and we have assumed an electron temperature of 10$^4$K and density in the range $10^2<N_e<10^4$\,cm$^{-3}$. The resulting gas mass in solar masses is:

\begin{equation}
 M_{\rm HII}=3\times10^{19}\left(\frac{F_{\rm Br\gamma}}{\rm erg\,cm^{-2}\,s^{-1}}\right)\left(\frac{D}{\rm Mpc}\right)^2\left(\frac{N_e}{\rm cm^{-3}}\right)^{-1}
\end{equation}

The integrated Br$\gamma$ flux  over the NLR is  $\approx 4.2\times10^{-14}\,{\rm erg\,s^{-1}\,cm^{-2}}$. As discussed above, there is an average reddening of E(B-V)=0.5 along the NLR. Correcting for this reddening, using the law of \citet{ccm89} the Br$\gamma$ flux is $F_{\rm Br\gamma}\approx 4.4\times10^{-14}\,{\rm erg\,s^{-1}\,cm^{-2}}$.
Adopting an electronic density of 100\,cm$^{3}$ we obtain $M_{HII}\approx 2.4\times10^6\,{\rm M_\odot}$. The adopted density is justified by the fact that the resulting H{\sc\,ii} mass, divided by the volume of two cones with height of 100\,pc and opening angle of 75$\degr$ -- the approximate geometry of the observed intensity distributions -- does result in a density value of $\approx$100\,cm$^{-3}$. It may be that the density is higher, and the filling factor is smaller than one, but the resulting mass will be the same.

We can also calculate the mass of hot H$_2$ (which emit the K-band emission lines) as in \citet{riffel08} and \citet{scoville82}:

\[
 M_{H_2}=\frac{2m_p\,F_{H_{2}\lambda2.1218}\,4\pi D^2}{f_{\nu=1,J=3}A_{S(1)}\,h\nu}
\]
\begin{equation} 
~~~~~~~~=5.0776\times10^{13}\left(\frac{F_{H_{2}\lambda2.1218}}{\rm erg\,s^{-1}\,cm^{-2}}\right)\left(\frac{D}{\rm Mpc}\right)^2,
\label{mh2}
\end{equation}

\noindent
where $m_p$ is the proton mass, $F_{H_{2}\lambda2.1218}$ is the line flux, $D$ is the distance to the galaxy and M$_{H_2}$ is given in solar masses. For a typical vibrational temperature of $T_{vib}=2000$\,K (similar to the value we have obtained) the population fraction  is $f_{\nu=1,J=3}=1.22\times10^{-2}$ and the  transition probability  is $A_{S(1)}=3.47\times10^{-7}$\,s$^{-1}$ \citep{riffel08,turner77,scoville82}. 

The total H$_2$\,2.1218 flux, integrated over the whole emitting region is $2.5\times10^{-14}$\,erg\,s$^{-1}$\,cm$^{-2}$. Correcting for the average reddening of E(B-V)=0.5,  we obtain $F_{H_2\lambda2.1218}=2.64\times10^{-14}$\,erg\,s$^{-1}$\,cm$^{-2}$. Then, using the expressions above we obtain M$_{H_2}$=240\,M$_\odot$.

This value is 10$^4$ times smaller than that of H{\sc\,ii}, but it should be noted that this mass is only of the hot  H$_2$, which emits  because either shocks (probably from the accretion flow along the bar) or X-rays from the AGN excite the H$_2$ molecule. Most of the molecular gas in the nuclear region of galaxies is cold, with  the hot-to-cold mass ratio ranging between 10$^{-7}$ to 10$^{-5}$ \citep{dale05}. Thus, the total mass (hot plus cold) of molecular gas is probably even larger than that of H{\sc \,ii}. The presence of such a molecular gas reservoir around AGN is supported by radio observations of the central region of active galaxies, by, for example, the ``NUGA'' group, which report molecular gas masses in the range 10$^7$--10$^9\,{\rm M_\odot}$ in the inner few hundred parsecs of active galaxies \citep[][e.g.]{boone07,gb05,krips07}.  In the case of NGC\,4151,  as discussed above,  \citet{mundell99a} have measured inflows in radio observations of H\,{\sc i} along the large scale bar (at PA 130$\degr$), which may be the origin of a molecular gas reservoir accumulated around the nucleus, only a small part of which we observe, due to excitation of the H$_2$ molecule.

\section{Summary and Conclusions}

We have mapped the emitting gas intensity distributions, reddening and excitation in the NLR of NGC\,4151 using Gemini NIFS observations of the inner $\approx 200 \times$300\,pc$^2$ of the galaxy, covering the near-IR Z, J, H and K spectral bands, at a resolving power R$\ge$5000 and spatial resolution of $\approx$8\,pc. The main results of this paper are:
\begin{itemize}
\item
The intensity distributions in the recombination lines of H and He as well as in  the  [S{\sc\,iii}],  [P{\sc\,ii}], [ [Fe{\sc\,ii}] and [S{\sc\,iii}] emission lines, similarly to that in the optical  [O{\sc\,iii}], are most extended along PA=$60/240\degr$ -- the axis of the previously known bicone. The emitting region is somewhat brighter and reaches a larger projected distance of $\approx$130\,pc from the nucleus to the SW (the near side of the bicone) than to the NE (the far side of the bicone), where it reaches a distance of $\approx$100\,pc. The fluxes in the recombination lines decrease with distance $r$ from the nucleus as $\propto\,r^{-1}$, while those on the other emission lines above  remain constant or even increase with distance from the nucleus along the NLR.

\item
The H$_2$ intensity distribution is completely different from that of the ionized gas, avoiding the region of the bicone, probably due to the destruction of the H$_2$ molecule by the strong ionizing flux along the bicone. Most of the H$_2$ emission seems to be coming not from the outflowing gas, but from gas which is in the plane of the galaxy. This is supported by its kinematics (Paper II). The concentration of the  H$_2$ emission approximately perpendicular to the bicone indicates that there is attenuation of the ionizing flux at these locations, precluding the destruction of the H$_2$ molecule. This attenuation may be due to an obscuring torus and/or to the bottom part of the biconical outflow (as suggested by \citet{kra08}). 


\item
The intensity distribution in the coronal line [Si{\sc\,vii}] is resolved and that in the [Ca{\sc\,viii}] is marginally resolved, consistent with an origin  in the inner NLR.  The  intensity distributions in these and the other coronal lines --  [S{\sc\,viii}] and [S{\sc\,ix}] -- decrease steeply with distance as $\propto\,r^{-2}$, similarly to those of the calibration stars. 
\item
The line ratios [Fe{\sc\,ii}]$\lambda$1.257/1.644 and Pa$\beta$/Br$\gamma$ were used to map the reddening along the NLR, which ranges from E(B-V)=0 to E(B-V)=0.5. Within 0$\farcs$5 from the nucleus, the reddening increases up to E(B-V)$\ge$1.

\item
The [S{\sc\,iii}]/Pa\,$\beta$ line ratio has a value of $\approx$11 to the SW (typical for the NLR of Seyfert galaxies), where we are looking at the inner wall of the near cone, and is $\approx$ 30\% lower to the NE, where we are looking at the outer wall of the far cone. This difference seems not to be due to reddening and  indicates  lower excitation to the NE. 

\item
The line-ratio map [Fe{\sc\,ii}]\,1.257/[P{\sc\,ii}]\,1.187 of the NGC\,4151 NLR is the first 2D such map of an extragalactic source, and, similarly to  the [Fe{\sc\,ii}]\,1.257/Pa\,$\beta$ ratio shows larger values at the locations where the [Fe{\sc\,ii}] emission is enhanced at $\approx$\,1\arcsec\ from the nucleus. The increase in these line ratios maps the shocks produced by the radio jet in the NLR, which release the Fe{\sc\,ii} usually tied up in dust grains. This is confirmed by the correlation between the linhe-ratio maps and the outer parts of the radio map. From the many emission lines observed at the locations where the [Fe{\sc\,ii}] emission is enhanced, we have obtained the gas density, N$_e\,\approx$\,4000\,cm$^{-3}$ and temperature T$_e\,\approx$\,15000$\pm$5000K.

\item
From the fluxes of 10 H$_2$ emission lines, we have concluded that the H$_2$ emitting gas is in thermal equilibrium at the excitation temperature $T_{exc}=2155\,K$, ruling out any significant contribution from fluorescence to the excitation of the H$_2$ molecule. The thermal excitation may be due to X-rays from the AGN escaping perpendicular to the bicone axis or else to shocks produced by the accretion flow observed along the bar \citep{mundell95} when the gas reaches the nuclear region.


\item
We have calculated the mass of the  ionized and molecular gas, obtaining for the former $M_{HII}\approx\,2.4\,\times10^6\,{\rm M_\odot}$ and for the latter only M$_{H_2}\,\approx$\,240\,M$_\odot$. This small  mass is  nevertheless only that  of the ``hot skin''  of a probably much larger molecular gas mass.

\item
The distinct intensity distribution and smaller temperature (as well as distinct kinematics; see Paper II) of the H$_2$ emission when compared with those of the ionized gas, supports a distinct origin for the emitting gas.  The H$_2$ emission is probably a tracer of a large molecular gas reservoir,  built up from the accretion flow observed along the large scale bar, which may  be the source of fuel to the AGN. The H$_2$ emission can thus be considered a tracer of the {\it feeding} of the AGN, while the ionized gas emission, which maps the outflowing gas, is a tracer of the {\it feedback} from the AGN.

\end{itemize}

\section*{Acknowledgments}
We thank an anonymous referee for valuable suggestions which helped to improve the paper, as well as Dr. Irapuan Rodrigues for help with the IDL routine.
Based on observations obtained at the Gemini Observatory, which is
operated by the Association of Universities for Research in Astronomy,
Inc., under a cooperative agreement with the NSF on behalf of the
Gemini partnership: the National Science Foundation (United States),
the Science and Technology Facilities Council (United Kingdom), the
National Research Council (Canada), CONICYT (Chile), the Australian
Research Council (Australia),  Minist\'erio da Ci\^encia e Tecnologia (Brazil) and SECYT
(Argentina). This work has been partially supported by the Brazilian
institution CNPq.

{}   
\end{document}